\documentclass[
reprint,
superscriptaddress,
 amsmath,amssymb,
 aps,
floatfix,
]{revtex4-1}

\usepackage{graphicx}
\usepackage{dcolumn}
\usepackage{bm}

\usepackage{gensymb}
\usepackage{color}
\usepackage{lipsum}
\usepackage{siunitx}

\begin{document}

\preprint{APS/123-QED}

\title{Glass Polymorphism in TIP4P/2005 Water:  A Description Based on the Potential Energy Landscape Formalism}

\author{Philip H. Handle}
\affiliation{%
 Institute of Physical Chemistry,
 University of Innsbruck,
 Innrain 52c, A-6020 Innsbruck, Austria
}
\author{Francesco Sciortino}%
\affiliation{%
 Department of Physics,
 Sapienza -- University of Rome,
 Piazzale Aldo Moro 5, I-00185 Roma, Italy
}

\author{Nicolas Giovambattista}
\affiliation{%
 Department of Physics,
 Brooklyn College of the City University of New York,
 New York, New York 10016, USA}
\affiliation{ Ph.D. Programs in Chemistry and Physics,
The Graduate Center of the City University of New York,
New York, New York 10016, USA
}

\date{\today}

\begin{abstract}
The potential energy landscape (PEL) formalism is a statistical mechanical approach to describe supercooled liquids and glasses.
Here we use the PEL formalism to study the pressure-induced transformations between low-density amorphous ice (LDA) and high-density amorphous ice (HDA) using computer simulations of the TIP4P/2005 molecular model of water.
We find that the properties of the PEL sampled by the system during the LDA-HDA transformation exhibit anomalous behavior.
In particular, at conditions where the change in density during the LDA-HDA transformation is approximately discontinuous, reminiscent of a  first-order phase transition,
we find that (i) the inherent structure (IS) energy, $e_\text{IS}(V)$, is a concave function of the volume, {\it and} (ii) the IS pressure, $P_\text{IS}(V)$, exhibits a van der Waals-like loop.  In addition, the curvature of the PEL at the IS is anomalous, a non-monotonic function of $V$.
In agreement with previous studies, our work suggests that conditions (i) and (ii) are necessary (but not sufficient) signatures of the PEL for the LDA-HDA transformation to be reminiscent of a first-order phase transition.
We also find that one can identify two different regions of the PEL, one associated to LDA and another to HDA. 
Our computer simulations are performed using a wide range of compression/decompression and cooling rates.  In particular, our slowest cooling rate (0.01~K/ns) is within the experimental rates employed in hyperquenching experiments to produce LDA.  Interestingly, the LDA-HDA transformation pressure that we obtain at $T=80$~K and at different rates extrapolates remarkably well to the corresponding experimental pressure.

\end{abstract}

\pacs{Valid PACS appear here}
\maketitle


\section{Introduction}

Water is a prototypical complex substance.
This complexity is manifested in numerous anomalous properties present in the liquid state~\cite{errington2001relationship,ludwig01-angew,debenedetti03-jpcm}, such as the maximum in density upon isobaric cooling (\SI{4}{\celsius} at 1~bar), and the maximum in diffusivity upon isothermal compression ($\approx200$~MPa at \SI{0}{\celsius}).
In the solid state, water can exist in a surprisingly large number of crystalline polymorphs (17 distinct ices have been identified so far~\cite{salzmann19-jcp}) and in at least two different glassy states (amorphous ices)~\cite{mishima98-nature, angell04-arpc, loerting06-jpcm, loerting11-pccp,handle2017supercooled}.
The most common forms of glassy water are low- (LDA) and high-density (HDA) amorphous ice.
Amorphous ices can be obtained utilizing several thermodynamic paths~\cite{burton35-nature, bruggeller80-nature, mishima84-nature, handle15-pccp} and the behavior of these systems is well documented~\cite{mishima85-nature,mishima94-jcp, loerting01-pccp, gromnitskaya01-prb, klotz05-prl, loerting06-prl, winkel08-jcp, handle2018experimental1, handle2018experimental2}.
Remarkably, most experimental studies indicate that, when properly annealed~\cite{perakis2017diffusive,handle2018experimental1}, LDA and HDA can be interconverted by sharp and reversible transformations reminiscent of first-order phase transitions between equilibrium states~\cite{mishima85-nature,mishima94-jcp, gromnitskaya01-prb, klotz05-prl, koza05-prl, winkel08-jcp, handle2018experimental1}.

The puzzling behavior of water has spawned several potential theoretical scenarios~\cite{angell08-science,gallo16-crev,handle2017supercooled,anisimov18-prx}.
According to the liquid-liquid critical point (LLCP) scenario, water is hypothesized to exist in two different liquid states at low temperature, a low- (LDL) and a high-density (HDL) liquid.
In addition, LDL and HDL are separated by a first-order phase transition line ending at an LLCP at higher temperatures ~\cite{poole92-nature}.
One relevant advantage of the LLCP scenario, relative to other available theoretical explanations~\cite{angell08-science,gallo16-crev,handle2017supercooled,anisimov18-prx}, is that the LLCP scenario naturally rationalizes the experimental phenomenology present in the amorphous ices.
Specifically, in the LLCP scenario, LDA and HDA are the glass counterparts of LDL and HDL, respectively.
Accordingly, the sharp LDA$\rightleftarrows$HDA transformation is a result of extending the liquid-liquid phase transition into the glass domain~\cite{poole92-nature,mishima98-nature}, 
explaining the sharpness of the LDA$\rightleftarrows$HDA transformation found in experiments.
Experimental evidence for the connection between LDA and LDL, and between HDA and HDL, can be found in Refs.~\onlinecite{mcmillan65-nature, johari87-nature, smith99-nature, handle12-prl, amann-winkel13-pnas, perakis2017diffusive}. 
However, the true nature of LDA and HDA~\cite{tse99-nature,johari00-pccp} and their relationship with the liquid state are still a matter of debate~\cite{johari14-tca, johari15-tca, stern15-tca, handle16-tca, shephard16-jpcl,fuentes19-prx,stern19-pnas}. 

In this work, we study the LDA$\rightleftarrows$HDA transformation in water using molecular dynamics (MD) simulations in conjunction with the potential energy landscape (PEL) approach~\cite{goldstein1969viscous,stillinger1982hidden,debenedetti01-nature, sciortino05-jsm,bookstillinger}.
The PEL approach is a powerful theoretical framework within statistical mechanics that has been used extensively to study the dynamic and thermodynamic behavior of liquids at low temperatures~\cite{sastry2001relationship,mossa2002dynamics,la2006relation,heuer2008exploring}, including water~\cite{sciortino03-prl,lanave04-jpcb,handle18-jcp,handle18-mp}.  
In particular, it allows one to express the Helmholtz free energy $F(N,T,V)$ of a liquid [and hence, the corresponding equation of state (EOS)] in terms of statistical properties of the PEL.
In the case of water, the PEL formalism has been successfully applied to obtain the EOS for the SPC/E~\cite{spce} and TIP4P/2005~\cite{abascal05-jcp} water models~\cite{sciortino03-prl,handle18-jcp}. 
Such an EOS can be used to extrapolate the behavior of the supercooled liquid to low temperatures.
Interestingly, the PEL-EOS for TIP4P/2005 predicts the existence of an LLCP at $T_\text{c}=175$~K, $P_\text{c}=175$~MPa, and $\rho_\text{c}=0.997$~g/cm$^3$~\cite{handle18-jcp}, consistent with other predictions for this model~\cite{abascal10-jcp, sumi13-rscadv, singh16-jcp, biddle17-jcp}. 
In the case of SPC/E water, the LLCP is estimated to be located below the Kauzman temperature~\footnote{
Within the PEL formalism, the Kauzman temperature $T_\text{K}$ is defined as the temperature where the liquid has access to only one basin, i.e., $S_\text{conf}=0$~\cite{sciortino05-jsm}.
This means that below $T_\text{K}$ the system cannot undergo structural changes and hence, it cannot show a phase separation.
}
 and hence, it is not accessible to the liquid state~\cite{scala00-nature,scala00-pre,sciortino03-prl}.

Besides liquids, the PEL approach has also been applied to study  several atomic and molecular glasses~\cite{shell2003energy,shell2004thermodynamics,bookstillinger}.
In particular, it was used to study the LDA$\rightleftarrows$HDA transformation in SPC/E and ST2~\cite{stillinger74-jcp} water~\cite{giovambattista03-prl, giovambattista16-jcp, giovambattista17-jcp}.
In the case of SPC/E water, where an LLCP is not accessible~\cite{scala00-nature,scala00-pre,sciortino03-prl}, the changes in the PEL properties sampled by the system during the LDA$\rightarrow$HDA transformation are rather smooth and change monotonically with density~\cite{giovambattista03-prl}.
These are the expected results for normal glasses, such as for a system of soft-spheres~\cite{sun18-prl,sciortino05-jsm}.
Instead, in the case of ST2 water, where the LLCP is accessible~\cite{poole92-nature, poole05-jpcm, cuthbertson11-prl, liu12-jcp, palmer14-nature, smallenburg2015tuning, palmer18-jcp, palmer18-crev}, the PEL properties sampled by the system during the LDA$\rightarrow$HDA transformation exhibit anomalous behavior consistent with a first-order like phase transition between the two glass states~\cite{giovambattista16-jcp, giovambattista17-jcp}.
This conclusion was also supported by a PEL study of a water-like monatomic model that exhibits liquid and glass polymorphism~\cite{sun18-prl}.

One limitation of the ST2 water model to study glassy water is its inability to reproduce the structure of HDA~\cite{chiu13-jcp,chiu14-jcp}. 
Instead, the TIP4P/2005 water model reproduces relatively well the structure of LDA and HDA~\cite{wong15-jcp}.
Thus, it is a natural question whether TIP4P/2005 water also exhibits PEL anomalies during the pressure-induced LDA$\rightarrow$HDA transformation as found in ST2 water.
In this work we address this question and study the PEL of TIP4P/2005 water during the pressure-induced LDA$\rightleftarrows$HDA transformation.
We note that the TIP4P/2005 model is presently regarded as one of the most realistic (rigid) models to study liquid and crystalline water~\cite{vega11-pccp}.
Moreover, several studies~\cite{abascal10-jcp, wikfeldt2011spatially, sumi13-rscadv, yagasaki14-pre, russo14-natcomm, singh16-jcp, biddle17-jcp, handle18-jcp} are consistent with the presence of an LLCP in TIP4P/2005 water.
TIP4P/2005 water also displays an apparent first-order transition between LDA and HDA~\cite{wong15-jcp}, as observed experimentally.

A peculiar property of glasses is that their properties depend on the preparation process considered, i.e.,  glasses are history-dependent materials~\cite{kob-binder-book}.
This implies that the LDA$\rightleftarrows$HDA transformation can be sensitive to the cooling and compression rates employed as was shown in MD simulation studies~\cite{giovambattista03-prl,giovambattista16-jcp,giovambattista17-jcp}.
Accordingly, in this work we pay particular attention to the effects of cooling ($q_\text{c}$) and/or compression ($q_\text{P}$) rates. 
Specifically, we are able to reach, for the first time, cooling rates as slow as $q_\text{c}=0.01$~K/ns which is comparable to experimental cooling rates necessary to avoid crystallization in hyperquenching techniques~\cite{bruggeller80-nature, dubochet81-jmic, mayer82-nature, mayer85-jap,kohl00-pccp}.
The slowest compression rate  employed here ($q_\text{P}=0.1$~MPa/ns) expands beyond the slowest compression/decompression rates studied so far in MD simulations but it is still about three orders of magnitude faster than the fastest experimental rate we are aware of ($\SI{65}{\giga\pascal\per\second}=6.5\cdot10^{-5}$~MPa/ns)~\cite{chen11-pnas} and more than seven orders of magnitude faster than the rates commonly used experimentally ($\approx10^{-1}$--10$^{1}$~MPa/s)~\cite{mishima94-jcp,winkel08-jcp,handle2018experimental1}.

The structure of this work is as follows. In Sec.~\ref{sec:pel-liq}, we discuss briefly 
the PEL formalism and its application to the study of liquids.  
In Sec.~\ref{sec:sim} we describe the computer simulation details and the numerical  methods employed.
The results are presented in Sec.~\ref{sec:res} where we discuss the 
PEL properties of TIP4P/2005 water during the
preparation of LDA (Sec.~\ref{sec:lda}) and during the pressure-induced LDA$\rightleftarrows$HDA 
transformation (Sec.~\ref{sec:trans}).  
A summary and discussion is included in Sec.~\ref{sec:concl}.

\begin{table*}
\small
\caption{Combinations of compression/decompression temperature $T$, cooling-rate $q_\text{c}$, and compression/decompression rate $q_\text{p}$ studied here.
For the case $T=280$~K the liquid was equilibrated at 280~K and 1~bar before compression/decompression.
For the case $q_\text{c}=0.01$~K/ns some cooling runs were started from the liquid equilibrated at 200~K and others from 240~K (all at 1~bar).
For all other cases the liquid was initially equilibrated at 240~K and 1~bar.
The slowest compressions, at $q_\text{p}=0.1$~MPa/ns, were started from the configurations obtained at 400~MPa during the compression runs at $q_\text{p}=1$~MPa/ns (decompressions at the slowest rate considered are not included).} 
\begin{center}
\begin{tabular}{p{6cm}p{2.5cm}p{2.5cm}}
\hline
\multicolumn{1}{c}{Changing Parameter} &  \multicolumn{2}{c}{Constant Parameters} \\
\hline
$T=20$, 80, 160, 200, 280~K			& $q_\text{c}=30$~K/ns	& $q_\text{p}=300$~MPa/ns \\
$q_\text{c}=0.01$, 0.1, 1, 30, 100~K/ns		&$T=80$~K			& $q_\text{p}=10$~MPa/ns \\
$q_\text{p}=0.1$, 1, 10, 300, 1000~MPa/ns	&$T=80$~K			& $q_c=0.1$~K/ns \\
\hline
\end{tabular}
\end{center}
\label{tab:param}
\end{table*}

\section{The PEL formalism for Liquids}
\label{sec:pel-liq}

The PEL approach, as introduced by Stillinger and Weber~\cite{stillinger1982hidden}, is a powerful tool to describe the properties of liquids.
For a system of $N$ \emph{rigid} water molecules, the PEL is a hypersurface embedded  in a $(6N+1)$-dimensional space.
It is defined by the potential energy of the system, $U(\vec{r}^N,\phi^N, \theta^N, \psi^N)$, as function of the $3N$ coordinates of the molecules' center of mass $\vec{r}$, and the  
corresponding $3N$ Euler angles $\phi$, $\theta$, $\psi$. 
At any given time $t$, the system is represented by a single point on the PEL with coordinates given by the values of $\vec{r}^N(t),\phi^N(t), \theta^N(t), \psi^N(t)$.
It follows that, as time evolves, the representative point of the system describes a trajectory on the PEL, as it moves from one \emph{basin} of the PEL  to another. 
A basin is defined as the set of points of the PEL that lead to the same local minimum by potential energy minimization.
The local minimum associated to a given basin is called inherent structure (IS) and its associated energy is denoted $e_\text{IS}$. 
Depending on the temperature considered, the representative point of the system may or may not be able to overcome potential energy barriers separating different basins.
Accordingly, different regions of the PEL may be accessible to the system depending on the temperature considered.

In the PEL framework, the canonical partition function can be formulated as a one-dimensional integral~\cite{debenedetti01-nature,sciortino05-jsm}
\begin{equation}
Z(T,V)=\int\Omega(e_\text{IS})\text{d}e_\text{IS}~\text{e}^{-\beta F_\text{basin}(e_\text{IS},T,V)},
\label{eq:part}
\end{equation}
where $\Omega(e_\text{IS})\text{d}e_\text{IS}$ is the number of basins with IS energy between $e_\text{IS}$ and $e_\text{IS}+\text{d}e_\text{IS}$, $F_\text{basin}(e_\text{IS},T,V)$ 
 is the average basin free energy of basins with IS energy $e_\text{IS}$, $\beta=1/k_\text{B}T$ and $k_\text{B}$ is Boltzmann's constant.
All basins containing a significant amount of crystalline order are by definition excluded in the integration over phase space in Eq.~\ref{eq:part}~\cite{sciortino05-jsm}.
The basin free energy can further be written as
\begin{equation}
  F_\text{basin}(e_\text{IS},T,V)= e_\text{IS}+ F_\text{vib}(e_\text{IS},T,V),
\end{equation}
where  $F_\text{vib}$  accounts for the vibrational motion of the system around the IS with energy $e_\text{IS}$.
If we assume that the PEL around an IS can be approximated by a quadratic function (harmonic approximation), we can write
\begin{align}
&\beta F_\text{vib}(e_\text{IS},T,V)\approx\left<\sum_{i=1}^{6N-3}\ln\left(\beta\hbar\omega_i(e_\text{IS},V)\right)\right>_{e_\text{IS}}.
\end{align}
Here, the values $\omega_i(e_\text{IS},V)$ are the $6N-3$ normal mode frequencies and $\hbar$ is Planck's constant in its reduced form.  
To separate the $T$ and $e_\text{IS}$ dependence of $F_\text{vib}$, we write
\begin{equation}
 \beta F_\text{vib}(e_\text{IS},T,V)=(6N-3)\ln\left(\beta A_0\right)+\mathcal{S}(e_\text{IS},V),
  \label{eq:free-harm}
\end{equation}
where
\begin{equation}
 \mathcal{S}(e_\text{IS},V)=\left<\sum_{i=1}^{6N-3}\ln\left(\frac{\hbar\omega_i(e_\text{IS},V)}{A_0}\right)\right>_{e_\text{IS}}.
\label{eq:shape}
\end{equation}
The latter is called the basin shape function and it quantifies the average local curvature of the PEL around the IS.
In Eq.~\ref{eq:shape}, the average is taken over all IS with energy $e_\text{IS}$ and  $A_0=1$~ kJ/mol is a constant that ensures the arguments of the logarithm to have no units.

It can further be shown that the system free energy $F(N,T,V)$ can be
expressed as
\begin{equation}
 F= F_\text{IS}(E_\text{IS},T,V)+F_\text{vib}(E_\text{IS},T,V),
 \label{eq:free1}
\end{equation}
where
\begin{align}
 F_\text{IS}(E_\text{IS},T,V)=  
 E_\text{IS}(N,T,V)-TS_\text{conf}(E_\text{IS}).
 \label{eq:free2}
\end{align}
\noindent Here $E_\text{IS}=\left<e_\text{IS}(N,T,V)\right>$ is the average energy of the IS sampled by the system and $S_\text{conf}$ is the configurational entropy. 
The latter is defined as
\begin{equation}
 S_\text{conf}(E_\text{IS})\equiv k_\text{B}\ln[\Omega(E_\text{IS}) \text{d}E_\text{IS}].
\end{equation}

In summary, the PEL formalism allows the free energy of the system $F(N,T,V)$ to be expressed in terms of three basic properties of the PEL at constant $N,T,V$~\cite{sciortino05-jsm,sastry2001relationship,debenedetti01-nature}:
\begin{itemize}
\item[(i)] the average energy of the IS sampled by the system, i.e., $E_\text{IS}$;
\item[(ii)] the number of IS with energy between $E_\text{IS}$ and  $E_\text{IS}+\text{d}E_\text{IS}$, i.e., $\Omega(E_\text{IS})\text{d}E_\text{IS}$;
\item[(iii)]  the average curvature of the PEL at the IS, as quantified by
$\mathcal{S}$.
\end{itemize}

For systems at constant $N,T,V$, thermodynamic arguments show that the system is in stable or metastable equilibrium if and only if~\cite{stanleybook}
\begin{equation}
 \frac{\partial^2 F}{\partial V^2 } >0.
 \label{eq:stability}
\end{equation}
Here and in the following, the partial derivatives are evaluated at constant $T$ and $N$.
 Eq.~\ref{eq:stability} states that $F(N,T,V)$ must be a convex function of $V$ along an isotherm at constant $N$. 
Alternatively, since $P= - \partial F/ \partial V $, Eq.~\ref{eq:stability} can be rewritten as
\begin{equation}
 \frac{\partial P}{\partial V} < 0.
  \label{eq:p-stability}
\end{equation}
Eq.~\ref{eq:p-stability} implies that the isothermal compressibility of the system must be positive.
If Eqs.~\ref{eq:stability} or \ref{eq:p-stability} are violated then the 
system is unstable and exhibits a phase transition.

\begin{figure}[b!]
\begin{center}
\includegraphics[width=\columnwidth]{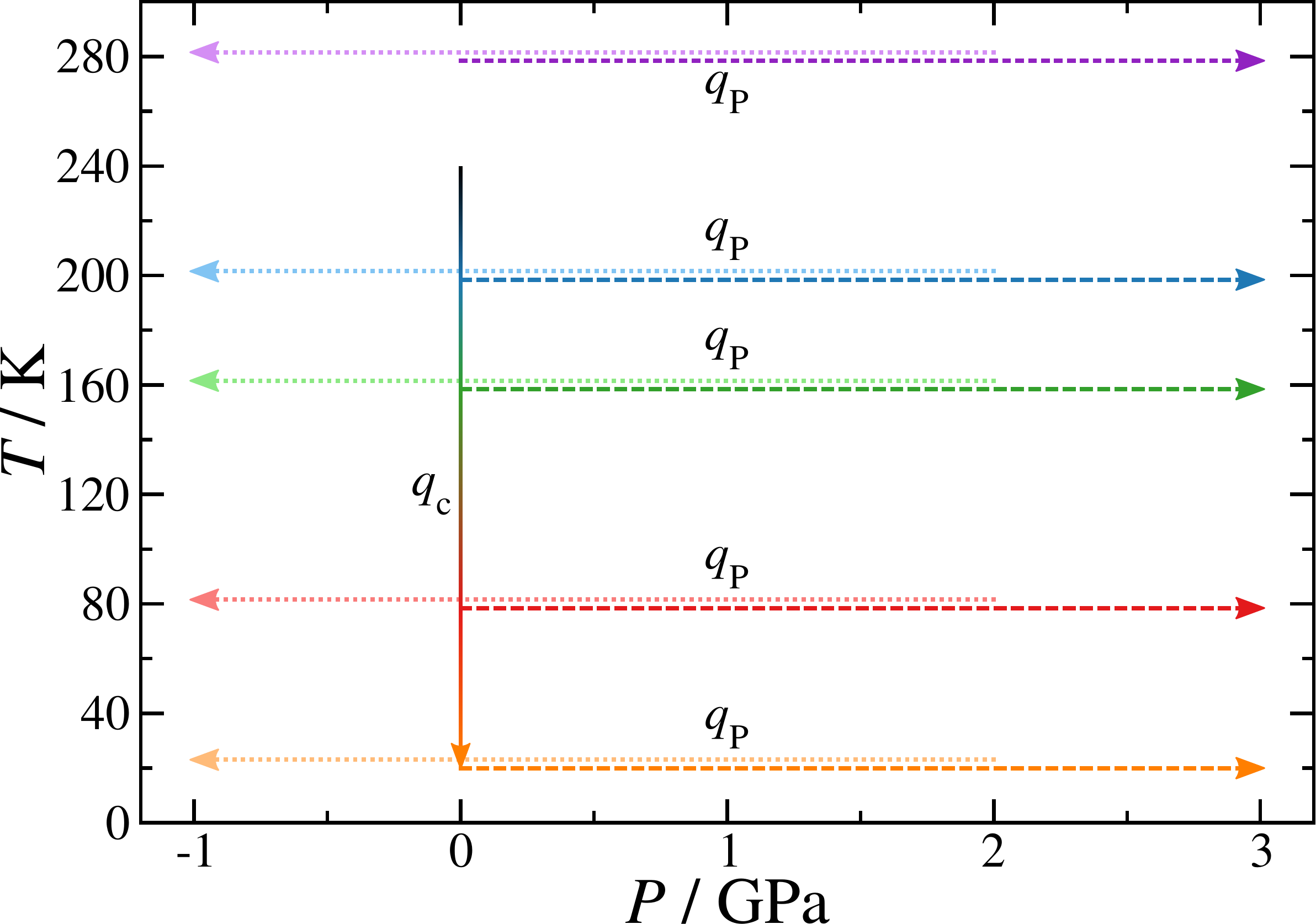}
\caption{
Schematic depiction of the $P$-$T$ paths taken in this study  (see also Table~\ref{tab:param}).
The liquid is equilibrated at $T_0=200$ or 240~K, and then cooled with rate $q_\text{c}$ to a target temperature $T$ (vertical arrow).
At this temperature the sample is isothermally compressed with rate $q_\text{P}$ up to 3~GPa (right arrows).
The configuration at 2~GPa is used as a starting configuration for the isothermal decompression with rate $q_\text{P}$ (left arrows).
The decompressions extend to negative pressures and end where the glass fractures.
Please note, that the compressions/decompressions at 280~K, in the liquid state, were performed starting from samples equilibrated at $T=280$~K and 1~bar.
}
\label{fig:pt-scheme}
\end{center}
\end{figure}

\begin{figure*}[t!]
\begin{center}
\includegraphics[width=0.70\textwidth]{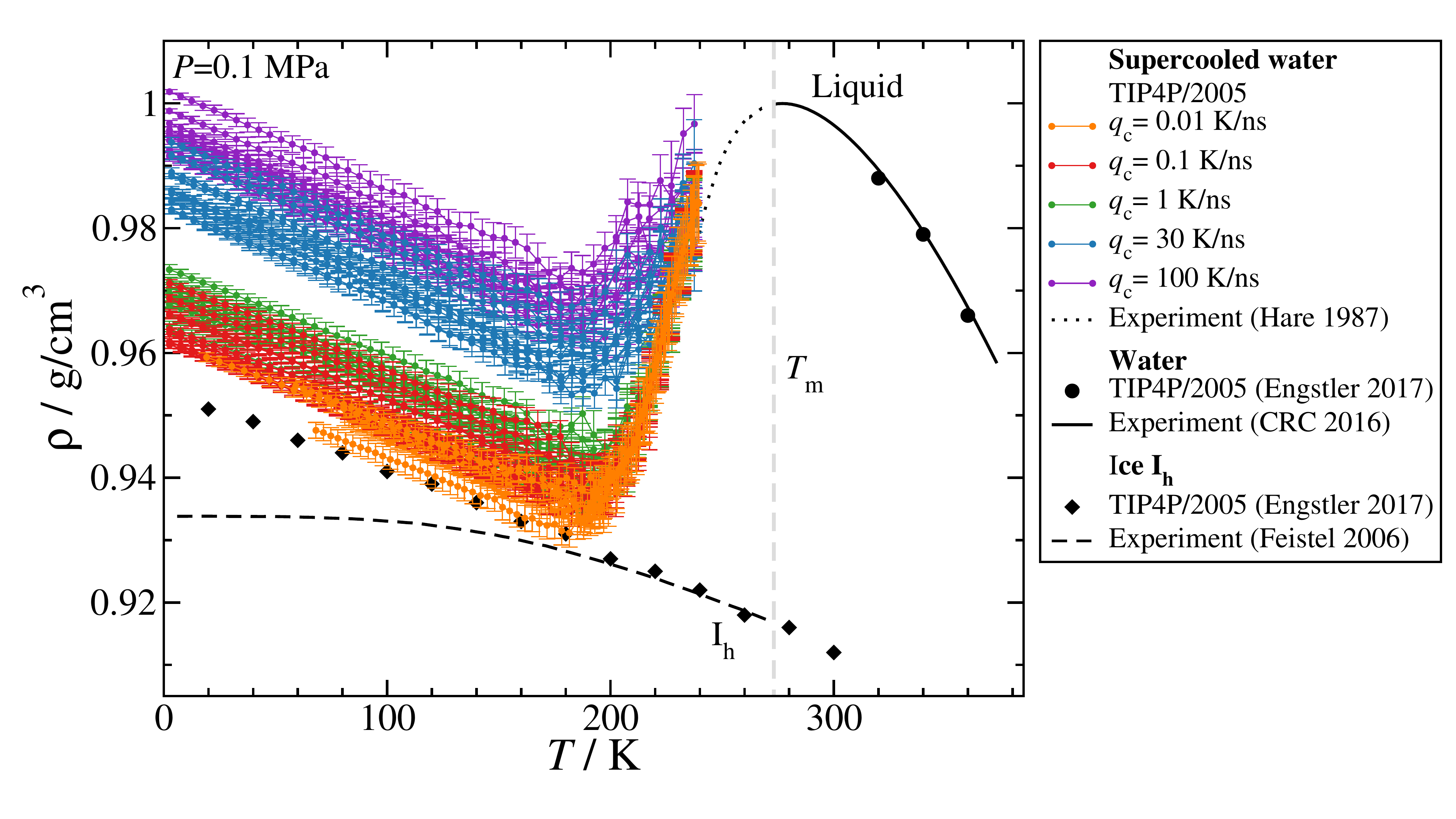}
\caption{Density as function of temperature upon cooling the liquid from $T_0=240$~K to the glass state (LDA) at different cooling rates $q_\text{c}$.
We show the results from all ten independent runs performed at each $q_\text{c}$.
For comparison we include the densities for TIP4P/2005 water in the hexagonal ice I (diamonds)~\cite{engstler17-jcp} and liquid state (solid circles)~\cite{engstler17-jcp}, as well as the experimental densities of ice I$_\text{h}$ (dashed line)~\cite{feistel06-jpcrd}, equilibrium liquid water (solid line)~\cite{crc2016}, and supercooled liquid water (dotted line)~\cite{hare87-jcp}. 
The slower $q_\text{c}$, the less dense is the final LDA state.
Deviation in the densities of ice I$_\text{h}$ from experiments and simulations are expected due to quantum nuclear effects.
$T_\text{m}$ is the experimental melting temperature of ice I$_\text{h}$.
}
\label{Fig:rho-vs-t}
\end{center}
\end{figure*}

Within the PEL formalism for liquids, Eq.~\ref{eq:stability} can be re-written in terms of Eq.~\ref{eq:free1} as
\begin{equation}
 \frac{\partial^2 F_\text{IS}}{\partial V^2}   +  \frac{\partial^2 F_\text{vib}}{\partial V^2} > 0. 
 \label{eq:is-stability}
\end{equation}
Similarly, Eq.~\ref{eq:p-stability} can be re-written as
\begin{equation}
\frac{\partial P_\text{IS}}{ \partial V} + \frac{P_\text{vib}}{ \partial V } < 0,  
 \label{eq:pis-stability}
\end{equation}
where $P=P_\text{IS} + P_\text{vib}$ and 
\begin{align}
P_\text{IS}&= - \frac{\partial F_\text{IS}}{\partial V},\label{eq:pis-stability2}\\
P_\text{vib}&= - \frac{\partial F_\text{vib}}{ \partial V}.
\end{align}
Eq.~\ref{eq:is-stability} shows that, for a system at constant $T$ and $N$, a phase transition may occur due to a concavity in $ F_\text{IS}(V)$, $F_\text{vib}(V)$, or both.
 Alternatively, Eq.~\ref{eq:pis-stability} implies that a phase transition can occur due to a positive slope in $P_\text{IS}(V)$,
  $P_\text{vib}(V)$, or both.

Eqs.~\ref{eq:is-stability} and \ref{eq:pis-stability} are thermodynamic stability conditions, in terms of PEL properties, that apply only to equilibrium systems.
Accordingly, they are not suitable to define phase transitions between glasses.
However, for the case of two different polyamorphic systems~\cite{giovambattista16-jcp,giovambattista17-jcp,sun18-prl}, it was shown that during the first-order-like phase transition between LDA and HDA,
$\partial^2 E_\text{IS}/ \partial V^2 >0$ and $\partial P_\text{IS}/ \partial V  <0$. 
For a system in equilibrium, this could lead to a violation of Eqs.~\ref{eq:is-stability} and \ref{eq:pis-stability}.
Indeed, if the basin shape function is approximately independent of the volume of the system then it can be shown that, within the harmonic approximation, $\partial P_\text{vib}/\partial V\approx 0$ and $\partial F_\text{vib}/\partial V\approx 0$.
In this case, the liquid exhibits a phase transition if and only if $\partial P_\text{IS}/ \partial V  <0$, or alternatively, if and only if $F_\text{IS}$ is a concave function of $V$ (at constant $N$ and $T$).

\section{Simulation Details}
\label{sec:sim}

The basis for our study of the LDA$\rightleftarrows$HDA transformation in TIP4P/2005 water are  MD simulations starting from LDA configurations.
We first prepare LDA at 1~bar by quenching the equilibrium liquid from temperature $T_0$ down to $T\leq80$~K, using cooling rates $q_\text{c}$ in the range $0.01$--100~K/ns (step (i); vertical arrow in Fig.~\ref{fig:pt-scheme}).
$T_0=240$~K for $q_\text{c}\geq 0.1$~K/ns and $T_0=200$ or 240~K for  $q_\text{c}= 0.01$~K/ns.
During this cooling from the liquid, we save 
configurations at different intermediate $T$. 
Configurations so prepared are then compressed up to $P=3$~GPa to produce HDA, using compression rates $q_\text{P}$ in the range 1--1000~MPa/ns (step (ii), horizontal right arrows in Fig.~\ref{fig:pt-scheme}).
Compressions at the slowest rate, $q_\text{P}=0.1$~MPa/ns, are started from $P=400$~MPa, using the respective configurations obtained during compression with rate $q_\text{P}=1$~MPa/ns as initial configurations.
HDA is then decompressed at the same rate starting from 2~GPa (step (iii), horizontal left arrows in Fig.~\ref{fig:pt-scheme}).

MD simulations to study glassy water have been criticized in the past due to the fast cooling and compression rates employed.
Accordingly, in this work, we explore in detail the influence of the compression/decompression temperature $T$ as well as rates $q_\text{c}$ and $q_\text{P}$ on the LDA$\rightleftarrows$HDA transformation observed in our MD simulations.
All combinations of $T$, $q_\text{c}$, and $q_\text{P}$ studied are listed in Table~\ref{tab:param}. 
We stress that the smallest cooling rate studied here ($q_\text{c}=0.01$~K/ns) is comparable to the rates used in hyperquenching experiments of liquid water~\cite{bruggeller80-nature, dubochet81-jmic, mayer82-nature, mayer85-jap,kohl00-pccp}.  With such a slow cooling rate,  we need to simulate  16~$\mu$s to cool the system from $240$ to $80$ K.

The systems studied consist of $N=1728$ TIP4P/2005 water molecules in a cubic box with periodic boundary conditions.
All our MD simulations are performed at constant $N$, $P$, $T$ using the GROMACS 5.1.4
and 2016.5~\cite{vanderspoel05-jcc} simulation packages.
Simulations use the leap-frog integrator with a time step of 2 fs.
Temperature is controlled using a Nos\'{e}-Hoover thermostat~\cite{nose84-mp, hoover85-pra} and pressure is controlled using a Parinello-Rahman barostat~\cite{parrinello81-jap}.
For the Coulomb interactions, we use a particle mesh Ewald treatment~\cite{essmann95-jcp} with a Fourier spacing of 0.1~nm.
For both the Lennard-Jones (LJ) and the real space Coulomb interactions, a cutoff of 0.85 nm is used.
Lennard-Jones interactions beyond 0.85 nm have been included assuming a uniform fluid density.
Water molecules are treated as rigid by using the LINCS (Linear Constraint Solver) algorithm~\cite{hess08-jctc} of 6th order with one iteration to correct for rotational lengthening.

At $T=240$~K the system was simulated for 10~ns.
After an initial equilibration period of 1~ns we extracted ten independent configurations separated by 1~ns.
These ten configurations served as the starting points for ten independent simulations [steps (i) to (iii)] for every set of compression/decompression temperature $T$ ($\leq200$~K),  $q_\text{c}$, and  $q_\text{p}$.
The length of these cooling and compression/decompression runs is determined by $q_\text{c}$ and $q_\text{P}$, respectively.

In order to calculate the properties of the PEL, i.e.,  $e_\text{IS}$, $P_\text{IS}$, ${\mathcal S}$, throughout the LDA$\rightleftarrows$HDA transformation, we obtain the IS during both the compression and decompression runs by minimizing the potential energy of the system every 10~MPa.
The minimization directly yields $e_\text{IS}$ and the Virial~\cite{frenkelbook} at the IS is used to calculate $P_\text{IS}$.
The basin shape function $\mathcal{S}$ is obtained from Eq.~\ref{eq:shape} using the normal mode frequencies given by the eigenvalues of the Hessian at the IS.

\section{Results}
\label{sec:res}
\subsection{Preparation of LDA}
\label{sec:lda}

\begin{figure}[b]
\begin{center}
\includegraphics[width=\columnwidth]{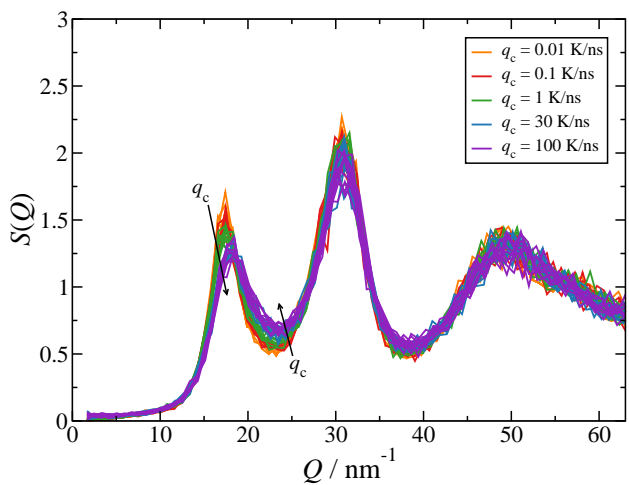}
\caption{Structure factor of LDA at 80~K and 1~bar prepared using different cooling rates $q_\text{c}$. For every $q_\text{c}$, all ten individual $S(Q)$ are shown.
}
\label{fig:sq-qdep}
\end{center}
\end{figure}

As discussed previously, 
to generate  LDA, the system is equilibrated at $T_0=240$~K and $P=1$~bar, and then cooled isobarically to different temperatures  $20\leq T\leq200$~K, using different cooling rates $q_\text{c}$.
The behavior of the density of the system upon cooling is shown in Fig.~\ref{Fig:rho-vs-t}.
In all cases the density first decreases, reaches a minimum around $200$~K and then increases linearly.
The influence of the cooling rate is clearly visible in Fig.~\ref{Fig:rho-vs-t}, which shows that the LDA samples obtained after cooling are less dense as $q_\text{c}$ is decreased.
As is typical for glasses, fast cooling increases the glass transition temperature, 
leaving the system trapped in higher density and higher energy regions of the PEL.
The very similar slope displayed by all samples in the low $T$ part (i.e., below the density minimum) results from the decrease in vibrations around the IS of the basins the systems are trapped in.

We note that the densities of the LDA forms obtained with $q_\text{c}=0.01$~K/ns (i.e., the experimental rate) are very similar to the density of TIP4P/2005 ice I$_\text{h}$.
This suggests that the LDA so produced consists of an almost fully developed, highly tetrahedral hydrogen-bond network.
Consistent with this assessment is the structure factor of LDA at 80~K reported in Fig.~\ref{fig:sq-qdep}.
It is visible that the pre-peak ($Q<20$~nm$^{-1}$) grows as $q_\text{c}$ is decreased.
At the same time it moves to lower $Q$ and separates more clearly from the main peak ($Q\approx30$~nm$^{-1}$).
This also indicates that lower $q_\text{c}$ yield an LDA with a more developed tetrahedral hydrogen-bond network.
The main peak in $S(Q)$ grows only slightly as $q_\text{c}$ is lowered and the features at higher $Q$ coincide within the noise of our data for all cooling rates studied.
The structural data in Fig.~\ref{fig:sq-qdep} indicate further, that our samples have not crystallized during cooling.
We confirm this by calculating the local order parameter as defined in Ref.~\onlinecite{russo14-natmat} and find that $>99\%$ of the molecules are classified as liquid (cf. also Refs.~\onlinecite{engstler17-jcp, martelli18-prm}).

\begin{figure}[t!]
\begin{center}
\includegraphics[width=\columnwidth]{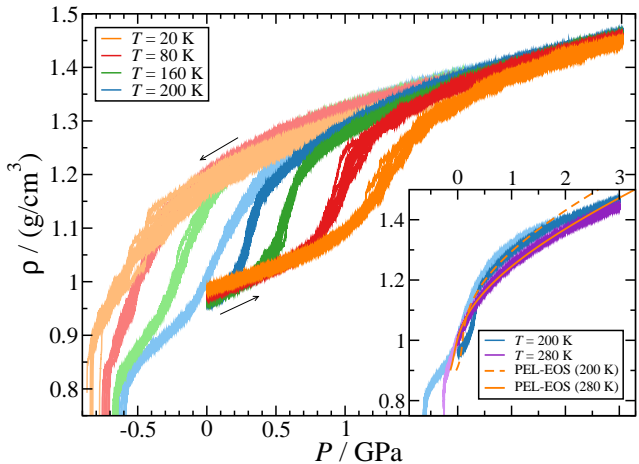}
\caption{
Density as function of pressure during the  compression of LDA and subsequent decompression of HDA (from $P=2$~GPa, lighter colors) at different temperatures.
LDA is prepared by isobaric cooling at $P=0.1$~MPa using a cooling rate $q_\text{c}=30$~K/ns.
The compression/decompression rate is $q_\text{P}=300$~MPa/ns.
Inset:  compression/decompression cycles at $T=200$ and 280~K, in the liquid state, compared to the 
corresponding prediction of the PEL-EOS from Ref.~\onlinecite{handle18-jcp} (solid and dashed orange lines).  Only at $T=280$~K, the system is in equilibrium at all pressures.
}
\label{fig:rho-tdep}
\end{center}
\end{figure}

\subsection{Pressure-Induced LDA-HDA Transformation}
\label{sec:trans}

Next, we discuss the properties of the system during the pressure induced LDA$\rightleftarrows$HDA transformation. 
In Sec.~\ref{sec:tdep}, we study the effects of varying $T$ at constant ($q_\text{c}$, $q_\text{P}$).
The effects of varying $q_\text{c}$ at constant ($q_\text{P}$, $T$), and $q_\text{P}$ at constant ($q_\text{c}$, $T$) are addressed in Secs.~\ref{sec:qdep} and \ref{sec:pidep}, respectively.

\subsubsection{Temperature Dependence}
\label{sec:tdep}

Samples of LDA obtained at $T=20$, 80, 160, 200, and 280~K and $P=0.1$~MPa, using a cooling rate $q_\text{c}=30$~K/ns, 
were compressed at constant temperature from $P=0.1$~MPa to 3~GPa with $q_\text{P}=300$~MPa/ns (see Fig.~\ref{fig:rho-tdep}).
At $T\leq 160$~K, i.e., below the estimated LLCP temperature $T_\text{c}\approx 175-193$~K~\cite{abascal10-jcp, sumi13-rscadv, singh16-jcp, biddle17-jcp, handle18-jcp},
 the system exhibits a sharp increase in density which signals the LDA$\rightarrow$HDA transformation. 
This density increase is sharper and shifts towards lower pressures as $T\rightarrow T_\text{c}$. 
 At 280~K ($T>T_\text{c}$), the system is in the liquid state and the density increases smoothly and monotonically during compression (inset, purple lines). 
 The behavior of $\rho(P)$  also overlaps with the corresponding equilibrium
 liquid isotherm obtained from the PEL-EOS in Ref.~\onlinecite{handle18-jcp} (inset, solid orange line).
At 200~K, however,  $\rho(P)$ (inset, blue lines) does not follow the corresponding PEL-EOS isotherm (inset, dashed orange line) and also shows a relatively sharp density step although no LDA$\rightarrow$HDA transformation is expected ($T>T_\text{c}$).
This indicates that the used compression rate is too large relative to the relaxation time of the liquid at 200~K and the system cannot reach equilibrium during compression.

\begin{figure}[b!]
\begin{center}
\includegraphics[width=\columnwidth]{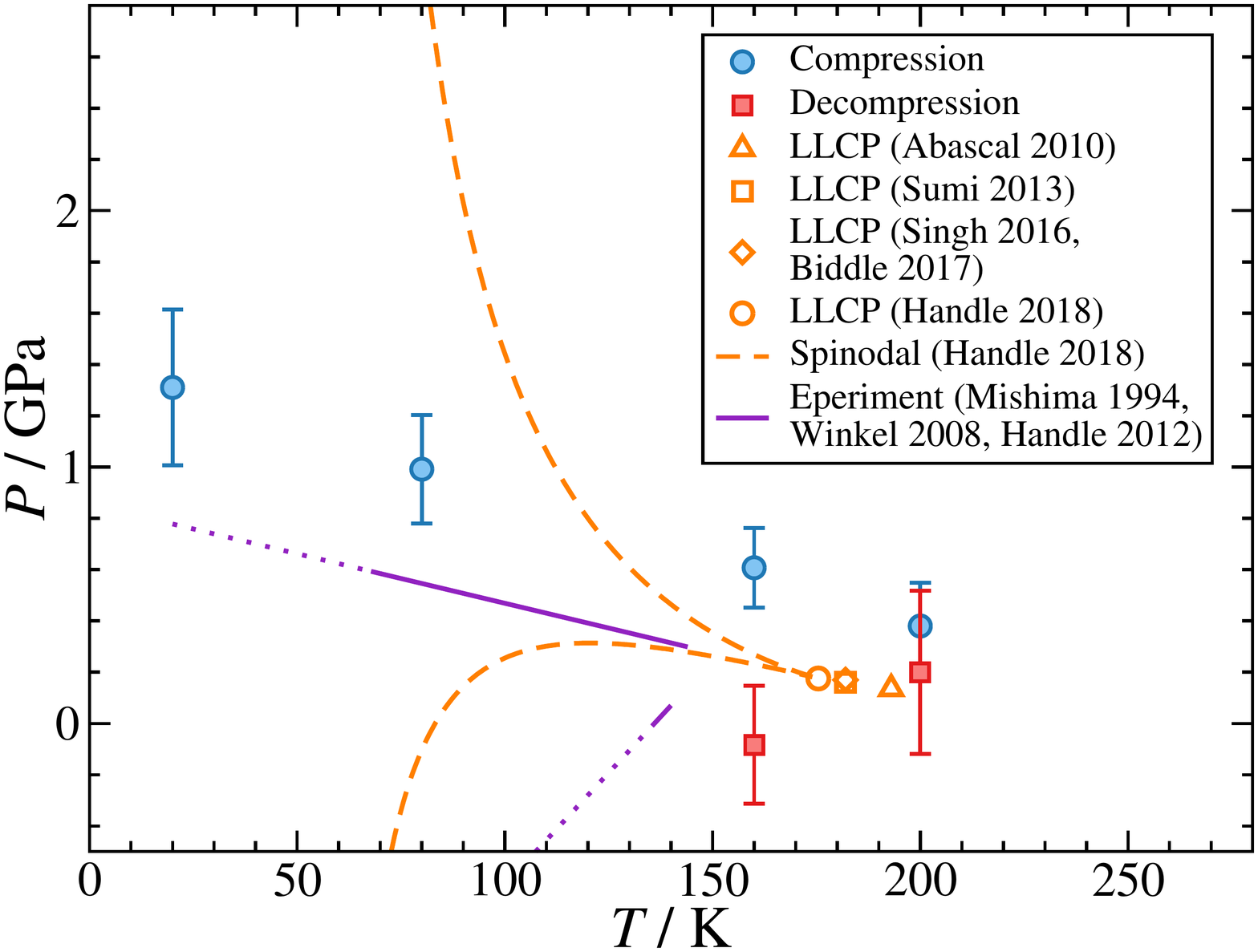}
\caption{
Pressure as function of temperature corresponding to the LDA$\rightarrow$HDA  (blue symbols) and HDA$\rightarrow$LDA (red symbols) transformations shown in Fig.~\ref{fig:rho-tdep}.
The (red and blue) symbols and associated `error bars' represent, respectively, the midpoint and 
the width of the corresponding transformation.
Orange symbols indicate the LLCP locations estimated in Refs.~\onlinecite{abascal10-jcp,sumi13-rscadv,singh16-jcp,biddle17-jcp,handle18-jcp}.
The orange dashed lines are the liquid-liquid spinodal lines based on the PEL-EOS of TIP4P/2005 water~\cite{handle18-jcp}.
The solid purple lines indicate the location of the LDA$\rightarrow$HDA and HDA$\rightarrow$LDA transformations from experiments~\cite{mishima94-jcp,winkel08-jcp,handle12-prl}.
The dotted lines are linear extrapolations of the experimental data to lower temperatures.
}
\label{fig:trans-tdep}
\end{center}
\end{figure}

The HDA configurations obtained at 2~GPa were used as the starting configurations for the decompression runs conducted
at the same $T$ and rate $q_\text{P}$ (see Fig.~\ref{fig:rho-tdep}).
At very low and negative pressures, all amorphous ices fracture.
This corresponds to the sudden (almost vertical) density drop at $P_\text{fract} <-500$~MPa in Fig.~\ref{fig:rho-tdep}.
Interestingly, $P_\text{fract}$ becomes more negative as $T$ decreases implying that the amorphous ices are stronger under tension at lower $T$.

The HDA$\rightarrow$LDA transformation, at $P_\text{fract}<P<0$, is very weak for the present model and at the studied rates. 
It is observable at 160~K ($T<T_\text{c}$) where $\rho(P)$ exhibits a change of slope at $\approx -85$~MPa and an LDA-like state can be identified at $-600< P<-300$~MPa.
However, at $T\leq80$~K, the HDA state seems to expand continuously until it finally fractures (cf. Refs~\onlinecite{wong15-jcp,engstler17-jcp}).
We also note that, a density step during the decompression of HDA is visible at $T=200$~K in Fig.~\ref{fig:rho-tdep}.
However, $\SI{200}{\kelvin}>T_\text{c}$ and hence, no HDA$\rightarrow$LDA transformation can exist at this temperature.
As mentioned previously, for the compression rate employed, the system is not able to constantly accommodate to the change in pressure at $T=200$~K, since the equilibration time is too long relative to the $q_\text{P}$ employed.
Accordingly, the densities during compression and decompression do not coincide with each other and also deviate from the equilibrium liquid isotherm obtained from the PEL-EOS (inset, dashed orange line)~\cite{handle18-jcp}. 
We note that the results shown in Fig.~\ref{fig:rho-tdep} are in full agreement with previous studies of glassy TIP4P/2005 water~\cite{wong15-jcp,engstler17-jcp} and in qualitative agreement with glassy ST2 water~\cite{poole92-nature, giovambattista16-jcp, giovambattista17-jcp}.

At  $T=280$~K the system is in the liquid state and $\rho(P)$ during the decompression coincides with the
density during compression.
In addition, $\rho(P)$ coincides with the density of the equilibrium liquid obtained from the PEL-EOS in Ref.~\onlinecite{handle18-jcp} (inset, solid orange line).

\begin{figure}
\begin{center}
\includegraphics[width=\columnwidth]{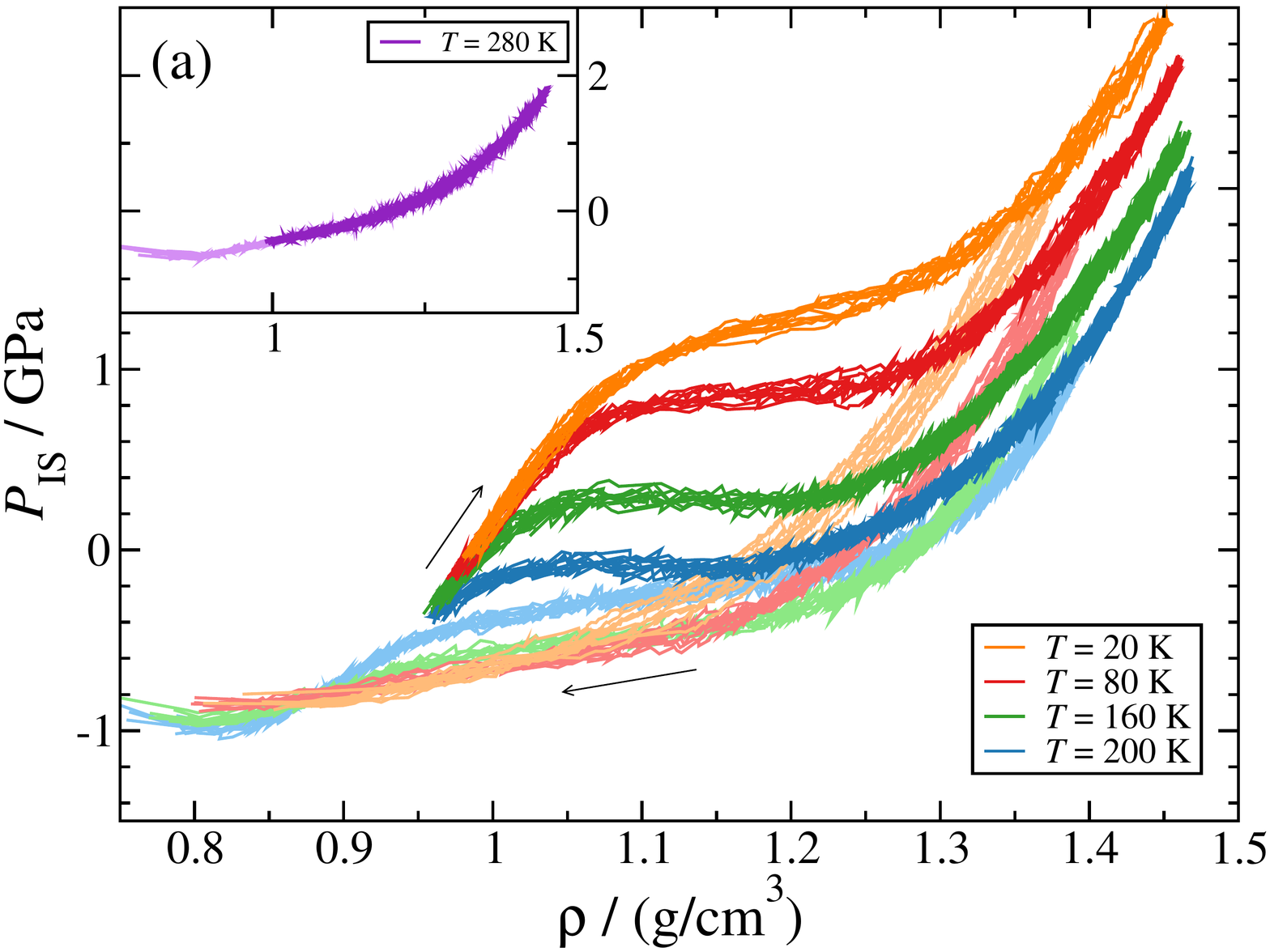}
\includegraphics[width=\columnwidth]{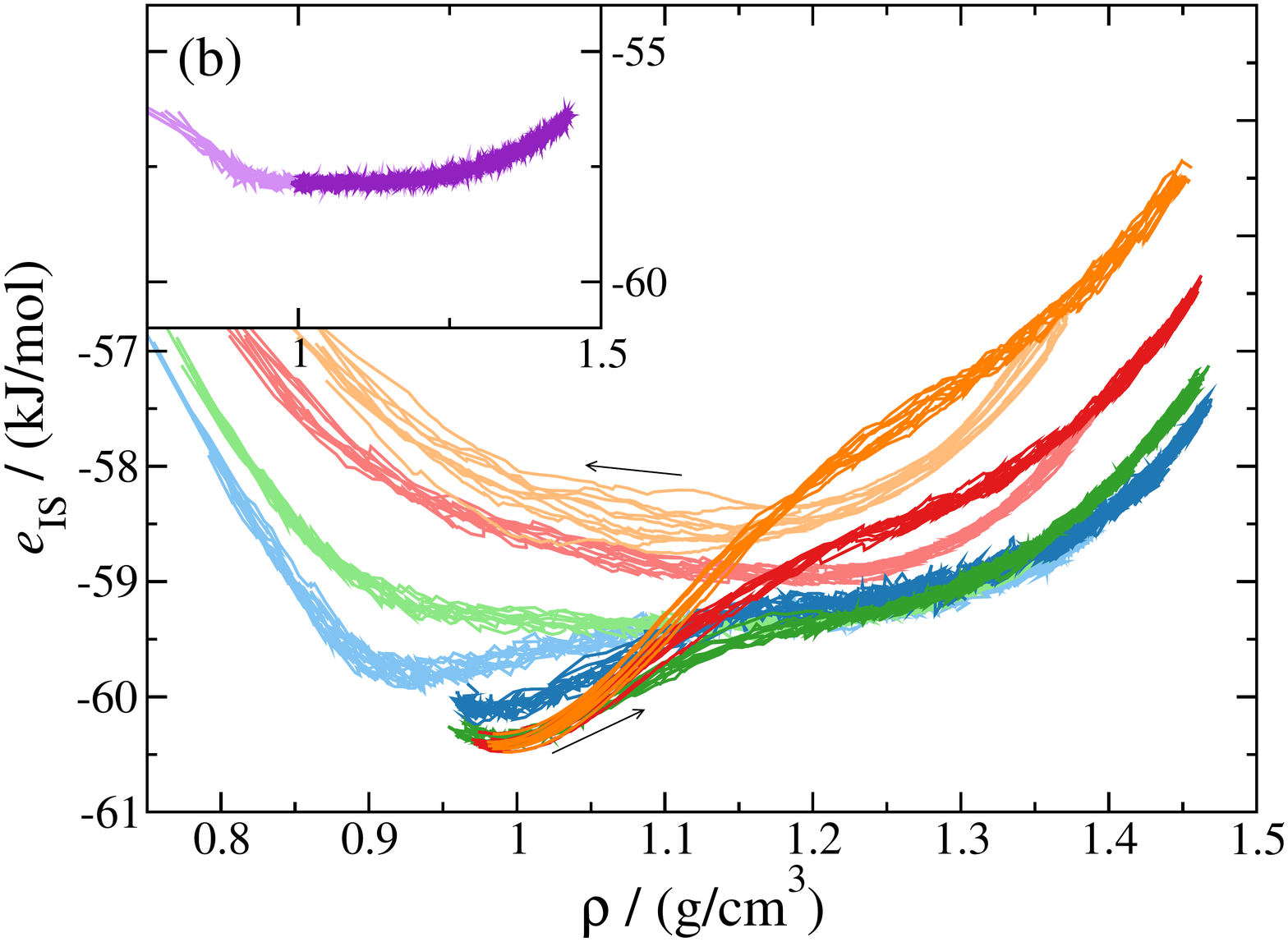}
\includegraphics[width=\columnwidth]{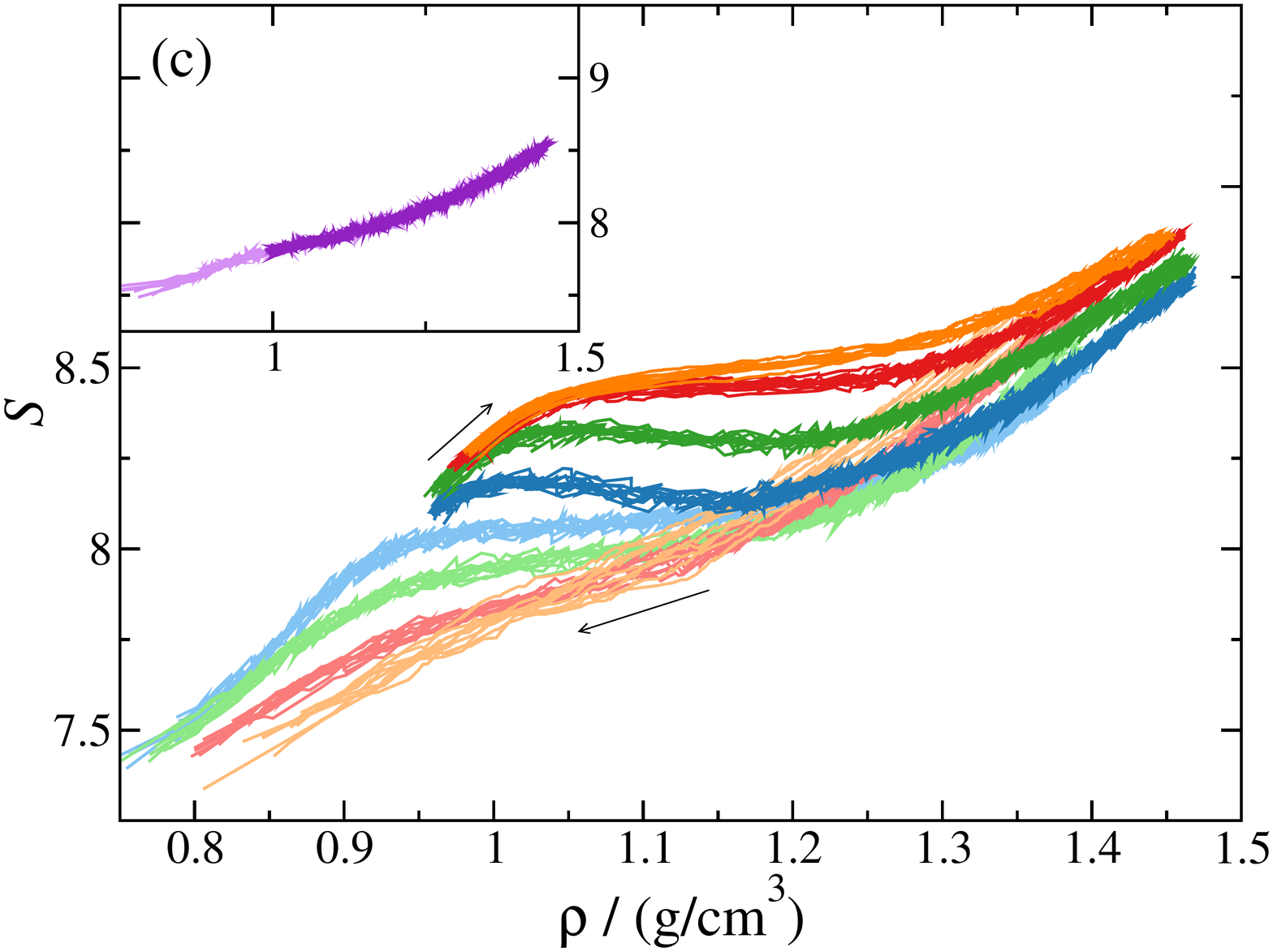}
\caption{
(a) Pressure, (b) energy, and (c) shape function of the IS sampled by the system during the compression/decompression cycles shown in Fig.~\ref{fig:rho-tdep}.
During compression, a weak  (anomalous) concavity region develops in $e_\text{IS}(\rho)$ at $T\leq200$~K and (anomalous) van der Waals-like loops are observable in $P_\text{IS}(\rho)$ and $\mathcal{S}(\rho)$ at $T=160$, 200~K.
The insets show $P_\text{IS}(\rho)$, $e_\text{IS}(\rho)$, and $\mathcal{S}(\rho)$ at $T=280$~K ($T>T_\text{c}$).
At this temperature the system reaches the equilibrium liquid state at all pressures, with no anomalies present in the PEL properties.
}
\label{fig:pel-tdep}
\end{center}
\end{figure}

Fig.~\ref{fig:trans-tdep} shows that the effect of increasing $T$ is to shift (i)  the LDA$\rightarrow$HDA transformation  (blue symbols) to lower pressures, and (ii) the HDA$\rightarrow$LDA (red symbols) transformation to higher pressures.
 This results in a narrower hysteresis during the LDA$\rightleftarrows$HDA transformation as $T\rightarrow T_\text{c}$.
 We stress that these results are qualitatively consistent with experiments (see purple lines).
 However, the LDA$\rightleftarrows$HDA transformation pressures obtained in our MD simulations (at the employed rates) are off relative to experimental values. 
 As we will show in Secs.~\ref{sec:qdep} and ~\ref{sec:pidep}, lowering $q_\text{c}$ and/or $q_\text{P}$ reduces
the hysteresis, improving the MD results relative to experiments.
We also note that the pressure-interval associated to the LDA$\rightarrow$HDA transformation (blue error bars) shrinks with increasing temperatures (i.e., the transformation becomes sharper).
 We expect a similar $T$-effect on the HDA$\rightarrow$LDA transformation.
 However, the HDA$\rightarrow$LDA transformation is not clearly observable in Fig.~\ref{fig:rho-tdep} at low temperatures.
 Accordingly, in Fig.~\ref{fig:trans-tdep}, the HDA$\rightarrow$LDA transformation is indicated only for $T =160$~K.
We also report the density steps for 200~K, although they are clearly not related to HDA$\rightleftarrows$LDA transformations.

Included in Fig.~\ref{fig:trans-tdep} are also the liquid spinodal lines associated to the LLCP  predicted by the PEL-EOS~\cite{handle18-jcp}.
Within the LLCP scenario, the LDA$\rightleftarrows$HDA transformations are nothing else but the extensions of the LDL$\rightleftarrows$HDL spinodal lines into the glass domain (cf., e.g., Ref.~\onlinecite{mishima98-nature}).
It follows from Fig.~\ref{fig:trans-tdep} that the liquid-liquid spinodal lines predicted by the PEL-EOS cannot be used to estimate 
 the LDA$\rightleftarrows$HDA transformation lines obtained either from MD simulations nor experiments (see purple lines).
This discrepancy is not due to a deficiency of the PEL formalism, but due to the fact that the PEL-EOS is parameterized based on IS sampled by the equilibrium liquid at $200<T<270$~K~\cite{handle18-jcp}.
As we show in the supplementary material, the IS sampled by the system during compression at low $T$ are not explored by the equilibrium liquid.
It is probably this difference between the PEL regions sampled by the liquid and the glass that makes the extension of the PEL-EOS into the glass domain of very limited applicability.

Next, we discuss the PEL properties of the system corresponding to the runs shown in Fig.~\ref{fig:rho-tdep}.
Fig.~\ref{fig:pel-tdep} shows $P_\text{IS}$, $e_\text{IS}$, and ${\mathcal S}$ as a function of density during the compression/decompression processes.
Consistent with studies based on the SPC/E and ST2 models~\cite{giovambattista03-prl,giovambattista16-jcp,giovambattista17-jcp}, we find that, at low temperatures, all PEL properties are different along the compression and decompression paths.
This implies that the system explores different regions of the PEL during the LDA$\rightarrow$HDA and HDA$\rightarrow$LDA transformations.
This is not the case for $T=280$~K, because, as explained previously, the system is in the equilibrium liquid state along both the compression and decompression processes.

The behavior of $P_\text{IS}(\rho)$  shown in Fig.~\ref{fig:pel-tdep}(a) is rather complex.
During compression at $T=280$~K, in the equilibrium liquid state, $P_\text{IS}(\rho)$ is a monotonic function of density, as expected. However, at $T=160$~K ($T<T_\text{c}$) and even at $T=200$~K ($T>T_\text{c}$), $P_\text{IS}(\rho)$ exhibits a van der Waals-like loop [i.e., a section of negative slope in $P_\text{IS}(\rho)$]. 
At these temperatures, the behavior of $P_\text{IS}(\rho)$ shown in Fig.~\ref{fig:pel-tdep}(a) is reminiscent of the behavior expected for equilibrium systems during a phase transition (cf. Sec.~\ref{sec:pel-liq}).
Interestingly, there is no van der Walls-like  loop at very low temperatures (80 and 20~K).

\begin{figure}[t!]
\begin{center}

\includegraphics[width=\columnwidth]{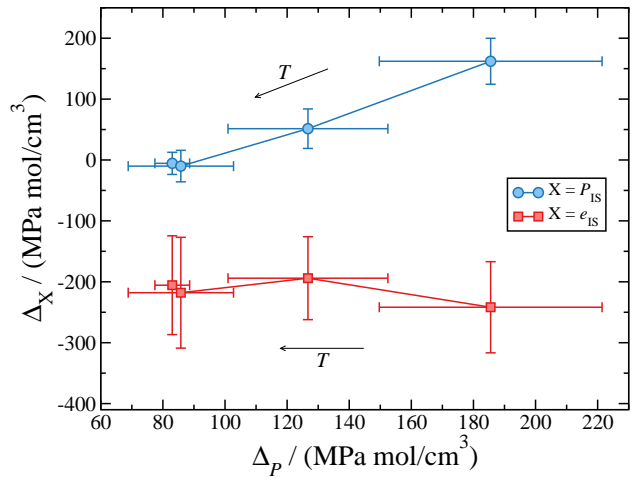}
\caption{
Relationship between the sharpness of the LDA$\rightarrow$HDA transformations shown in Fig.~\ref{fig:rho-tdep} and the corresponding anomalous character of the PEL properties (Fig.~\ref{fig:pel-tdep}).
$\Delta_P$ and  $\Delta_{P_\text{IS}}$ are the slopes of $P(\rho)$ and $P_\text{IS}(\rho)$ during the LDA$\rightarrow$HDA transformations (see Eqs.~\ref{eq:delta} and \ref{eq:delta-pis}).
$\Delta_{e_\text{IS}}$ quantifies the concavity in $e_\text{IS}$ during the transformation (see Eq.~\ref{eq:delta-eis}).
As $\Delta_P\rightarrow 0$ and the transformation becomes sharper (reminiscent of a first-order phase transition), $\Delta_{P_\text{IS}}$ decreases and becomes negative, i.e., $P_\text{IS}(\rho)$ develops an anomalous van der Waals-like loop (see Fig.~\ref{fig:pel-tdep}).
At the present cooling and compression rates, $\Delta_{e_\text{IS}}$ is negative, i.e., $e_\text{IS}(\rho)$ is anomalously concave, and remains roughly independent of the sharpness of the LDA$\rightarrow$HDA transformation.
}
\label{fig:delta-tdep}
\end{center}
\end{figure}

It follows from Figs.~\ref{fig:rho-tdep} and \ref{fig:trans-tdep} that, as $T$ increases from 20 to 160~K, the
 LDA$\rightarrow$HDA transformation becomes sharper, more reminiscent of a first-order phase transition, and accordingly,  $P_\text{IS}(\rho)$ develops a van der Waals-like loop [see Fig.~\ref{fig:pel-tdep}(a)]. 
To make this point clear, we compare the slopes of $P(\rho)$ and $P_\text{IS}(\rho)$ at the mid-point of the LDA$\rightarrow$HDA transformation in Fig.~\ref{fig:delta-tdep}.
We use the following notation (cf. Ref.~\onlinecite{giovambattista17-jcp}): 
\begin{equation}
\Delta_P=\left.-\frac{\partial P}{\partial V_\text{m}}\right|_{V_\text{m,mid}}
\label{eq:delta}
\end{equation}
 and
 \begin{equation}
\Delta_{P_\text{IS}}=\left.-\frac{\partial P_\text{IS}}{\partial V_\text{m}}\right|_{V_\text{m,mid}}.
\label{eq:delta-pis}
\end{equation}
Here $V_\text{m}$ denotes the molar volume and $V_\text{m,mid}$ denotes the molar volume at the midpoint of the transformation.
We note that $\Delta_P=0$ corresponds to a discontinuous density jump.
Hence, the closer $\Delta_P$ is to zero, the sharper  the LDA$\rightarrow$HDA transformation.
We also note that negative values for $\Delta_{P_\text{IS}}$ indicate a van der Waals-like loop in $P_\text{IS}(\rho)$.
It follows from Fig.~\ref{fig:delta-tdep} that, as $\Delta_P$ decreases, $\Delta_{P_\text{IS}}$ also decreases and becomes negative.
That is, the sharper the transformation, the more pronounced is the van der Walls-like  loop in $P_\text{IS}(\rho)$.
Consistent with Fig.~\ref{fig:rho-tdep}, we find that during the decompression runs,  $P_\text{IS}(\rho)$ is very smooth showing no van der Waals-like loop, at least for the rates considered [see Fig.~\ref{fig:pel-tdep}(a)].

The behavior of $e_\text{IS}(\rho)$ is shown in Fig.~\ref{fig:pel-tdep}(b).
During compression at $T=280$~K, the system is able to reach the equilibrium liquid state and hence, it samples the same  values of  $e_\text{IS}$ accessed by equilibrium liquid (see supplementary material).
However, at all lower temperatures, a concavity in  $e_\text{IS}$  develops during the LDA$\rightarrow$HDA transformation, 
a feature again reminiscent of the behavior expected for equilibrium systems during phase transitions (cf. Sec.~\ref{sec:pel-liq}).
To clarify this point, we show in Fig.~\ref{fig:delta-tdep} the minimum curvature (i.e., maximum concavity) of $e_\text{IS}(\rho)$ during the LDA$\rightarrow$HDA transformation, 
 \begin{equation}
\Delta_{e_\text{IS}}=\min\left[\frac{\partial^2 e_\text{IS}}{\partial V_\text{m}^2}\right],
\label{eq:delta-eis}
\end{equation}
as a function of $\Delta_P$.
Interestingly, $e_\text{IS}(\rho)$ exhibits a mild concavity at all $T<280$~K, even at 20 and 80 K where $P_\text{IS}(\rho)$ shows no van der Walls-like loop.
It follows that, a concavity in $e_\text{IS}(\rho)$ is {\it not} a sufficient (anomalous) property of the PEL for a glass to exhibit a first-order-like transition.
Indeed, as argued in Ref.~\onlinecite{sun18-prl}, the van der Waals-like loop in $P_\text{IS}(\rho)$ and a concavity in $e_\text{IS}(\rho)$ seem to be necessary (but not sufficient) conditions for a glass to exhibit a first-order-like phase transition.  
In addition, we note that the concavity in $e_\text{IS}(\rho)$ is rather $T$-independent (at least for $q_\text{c}=30$~K/ns and $q_\text{P}=300$~MPa/ns).  

During decompression, $e_\text{IS}(\rho)$ decreases monotonically with decreasing density until the system reaches its limit of stability at $\rho\approx  0.85-0.90$~g/cm$^3$, where the amorphous ices are prompt to fracture [see Fig.~\ref{fig:pel-tdep}(b)].
An exception is the case of $T=200$~K, were a very mild concavity seems to develop at $\approx1.1$~g/cm$^3$.
Interestingly, the amorphous ices obtained at $\approx0.9$~g/cm$^3$ after decompression have a very large IS energy relative to the starting LDA configurations at $\approx0.95$~g/cm$^3$.
This implies that the recovered LDA-like states are stressed glasses located in high regions of the PEL, within the corresponding LDA domain (cf. also Ref.~\onlinecite{giovambattista16-jcp}).

$\mathcal{S}(\rho)$ behaves similarly to $P_\text{IS}(\rho)$ [see Figs.~\ref{fig:pel-tdep}(a) and (c)].
Specifically, during compression at 160 and 200~K, $\mathcal{S}(\rho)$ shows a van der Waals-like loop.
In other words, the sampled basins become narrower during compression up to $\approx1.05$~g/cm$^3$, then become wider during compression up to $\approx1.2$~g/cm$^3$ and then become narrower again.
At 20 and  80~K (glass state), and at 280~K (liquid state), $\mathcal{S}(\rho)$ increases monotonically upon compression. 
During the decompression runs, $\mathcal{S}(\rho)$ decays monotonically with decreasing $\rho$ at all $T$ studied, i.e., the basins sampled by the system become wider as density decreases.
Fig.~\ref{fig:pel-tdep}(c) is fully consistent with previous studies on glassy ST2 water~\cite{giovambattista16-jcp,giovambattista17-jcp}.
We note, however, that in the case of a water-like monoatomic system that exhibits an LDA$\rightleftarrows$HDA transformation, $\mathcal{S}(\rho)$ shows no van der Waals-like loop during the LDA$\rightleftarrows$HDA transformation~\cite{sun18-prl}.

We conclude this section with a brief discussion on the phenomenology found for the compressions and decompressions at 200~K.
As noted above this temperature is higher than all $T_\text{c}$ estimates for TIP4P/2005~\cite{abascal10-jcp, sumi13-rscadv, singh16-jcp, biddle17-jcp, handle18-jcp}.
Hence, it may be surprising that we found relatively sharp density steps during compression and decompression (including a hysteresis), as well as a van der Waals-like loop in $P_\text{IS}$ and a concavity in $e_\text{IS}$.
These are signatures that we also find for the LDA$\rightarrow$HDA transformation, and which are similar to what is expected for first-order phase transitions.
Since $\SI{200}{\kelvin}>T_\text{c}$, a phase transition is ruled out as the cause of this phenomenology.
Instead, the anomalous properties of the PEL at $T=200$~K can be rationalized by noticing that the compression rate used is large enough so that the system is not able to reach equilibrium during compression and decompression.
This reminds us that phenomena observed in non-equilibrium systems should be interpreted with caution.
We note that, as we will show in Sec.~\ref{sec:pidep}, a decrease in the compression rate $q_\text{P}$ {\it increases} the sharpness 
of the LDA$\rightleftarrows$HDA transformation at 80~K (and leads to a more pronounced van der Waals-like loop in $P_\text{IS}$).
Instead, at $T=\SI{200}{\kelvin}>T_\text{c}$, reducing $q_\text{P}$ must bring the system to equilibrium, as we find for the case $T=280$~K.
Accordingly, a slower rate $q_\text{P}$ must {\it decrease} the sharpness of the apparent of the LDA/LDL-HDA/HDL transformation at $T=200$~K.

\subsubsection{Cooling-Rate Dependence}
\label{sec:qdep}

\begin{figure}[b!]
\begin{center}
\includegraphics[width=\columnwidth]{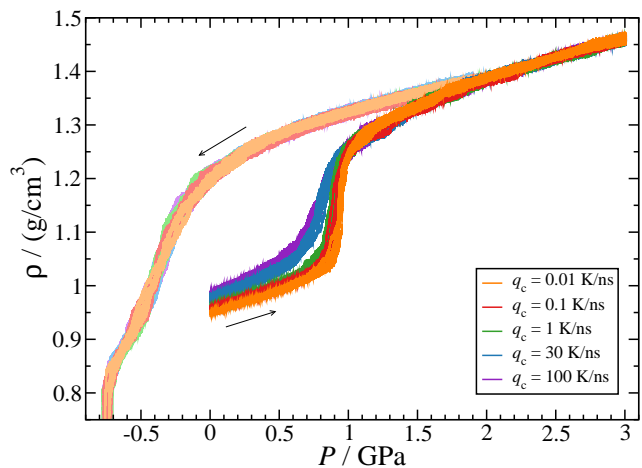}
\caption{
Density as function of pressure during the  compression of LDA and subsequent decompression of HDA (from $P=2$~GPa, lighter colors) at $T=80$~K.
LDA is prepared by isobaric cooling at $P=0.1$~MPa using using different cooling rates  $q_\text{c}=0.01$--100~K/ns.
The compression/decompression rate is $q_\text{P}=10$~MPa/ns.
}
\label{fig:rho-qdep}
\end{center}
\end{figure}

In order to study the cooling rate effects on the LDA$\rightleftarrows$HDA transformation, we consider
LDA samples prepared at 1~bar and 80~K using cooling rates $q_\text{c}=0.01$--100~K/ns.
All samples are then compressed/decompressed at $T=80$~K using $q_\text{P}=10$~MPa/ns.
Again, we point out that the smallest cooling rate studied here (0.01~K/ns~$=10^7$~K/s) corresponds to the estimated rate reached in experimental hyperquenching techniques~\cite{bruggeller80-nature, dubochet81-jmic, mayer82-nature, mayer85-jap,kohl00-pccp}.

\begin{figure}[t!]
\begin{center}
\includegraphics[width=\columnwidth]{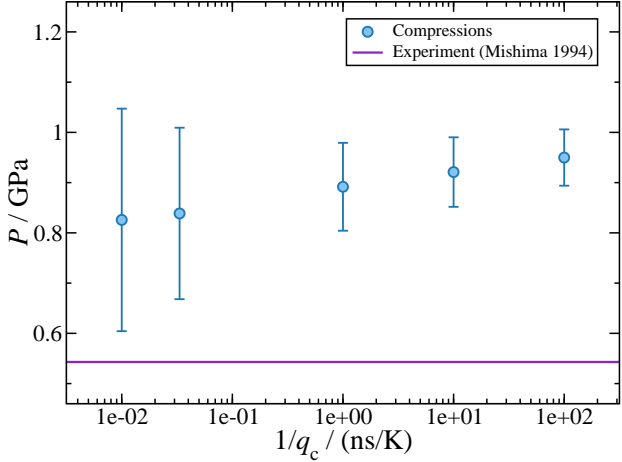}
\caption{
Pressure as function of the inverse of the cooling rate corresponding to the LDA$\rightarrow$HDA transformations shown in Fig.~\ref{fig:rho-qdep}.
The circles and associated `error bars' represent, respectively, the midpoint and the width of the corresponding transformation.
The solid purple line indicates the experimental LDA$\rightarrow$HDA transformation pressure obtained in the experiments of Ref.~\onlinecite{mishima94-jcp} (at much slower compression rate than those employed in the MD simulations).
}
\label{fig:trans-qdep}
\end{center}
\end{figure}

Fig.~\ref{fig:rho-qdep} reports $\rho(P)$ during the compressions/decompressions cycles starting from LDA forms prepared using different $q_\text{c}$.
In close similarity to our discussion in Sec.~\ref{sec:tdep}, Fig.~\ref{fig:rho-qdep} shows that all LDA samples experience a sudden densification that signals the LDA$\rightarrow$HDA transformation during compression. 
Instead, the HDA$\rightarrow$LDA transformation is rather smooth with no evident LDA-like state recovered at negative pressure.
HDA evolves continuously until it fractures at $P< -0.7$~GPa.

Fig.~\ref{fig:rho-qdep} clearly shows that, as $q_\text{c}$ decreases, the densification step associated to the LDA$\rightarrow$HDA transformation
becomes sharper and shifts slightly to higher $P$.
We stress that, the LDA$\rightarrow$HDA transformation at $q_\text{c}=0.1$--0.01~K/ns is remarkably similar to the experimental results~\cite{mishima94-jcp,handle2018experimental1}.
Indeed, based on the slope of the transformation, it is difficult to deny the first-order phase transition nature of the LDA$\rightarrow$HDA transformation.
Instead, during decompression of HDA, $\rho(P)$ decreases monotonically, with no evident LDA-like state (at the present conditions).
Interestingly, $\rho(P)$ behaves identically during decompression for all samples considered.  
This indicates that, once HDA forms, the system looses memory of the process followed to prepare LDA.
Accordingly, the HDA$\rightarrow$LDA transformation is unique for $T=80$~K and $q_\text{P}=10$~MPa/ns.  

\begin{figure}[t!]
\begin{center}
\includegraphics[width=0.965\columnwidth]{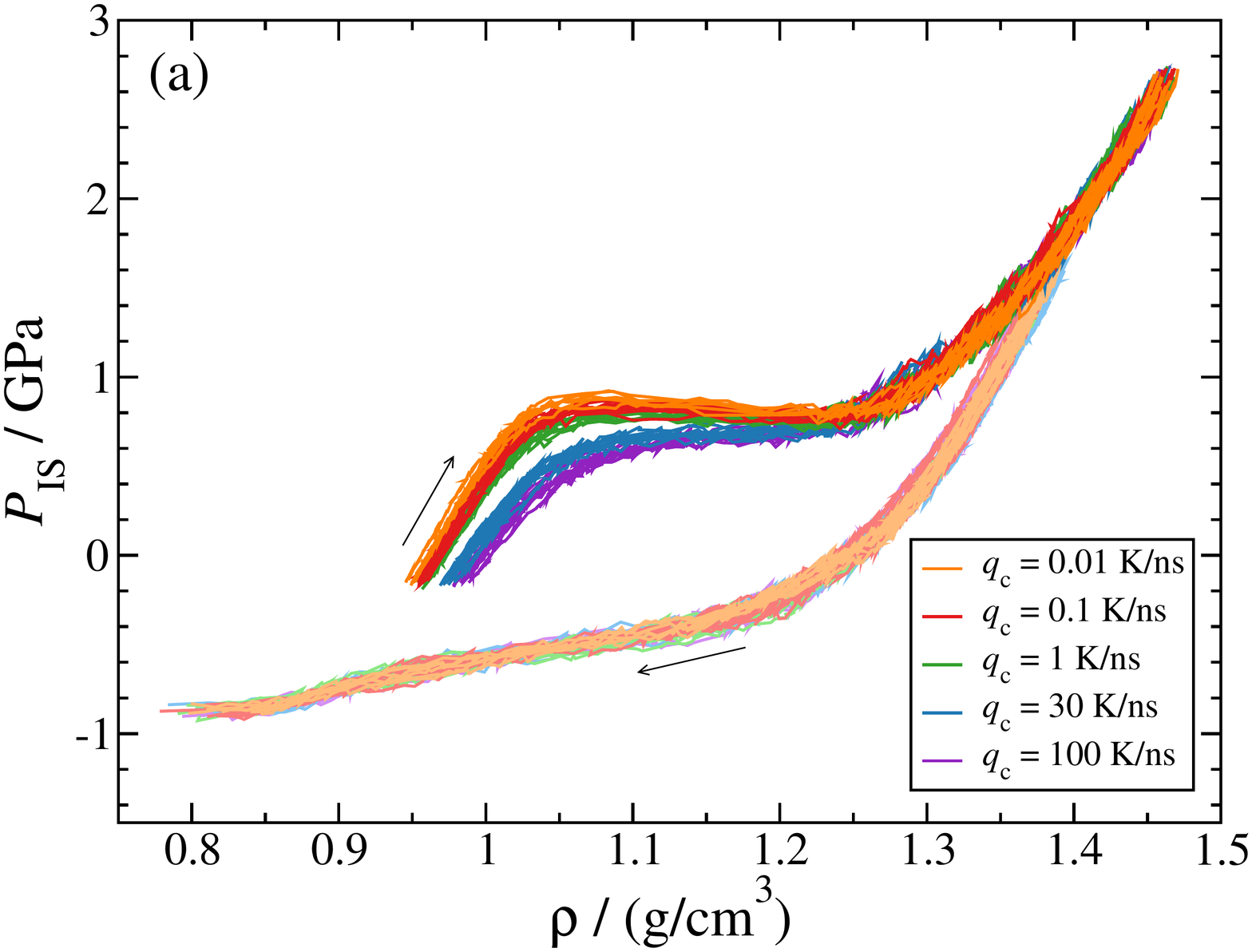}
\includegraphics[width=0.965\columnwidth]{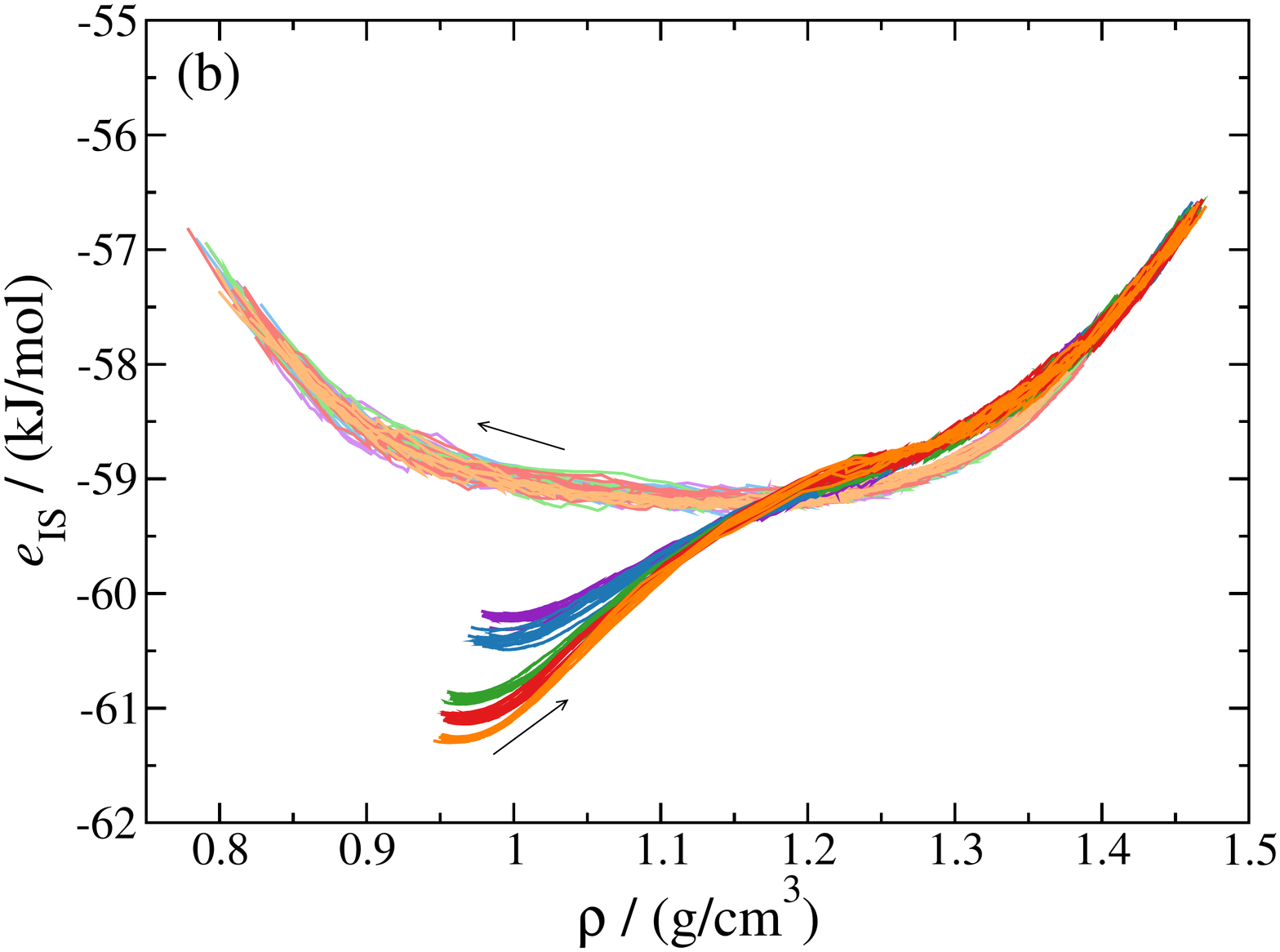}
\includegraphics[width=0.965\columnwidth]{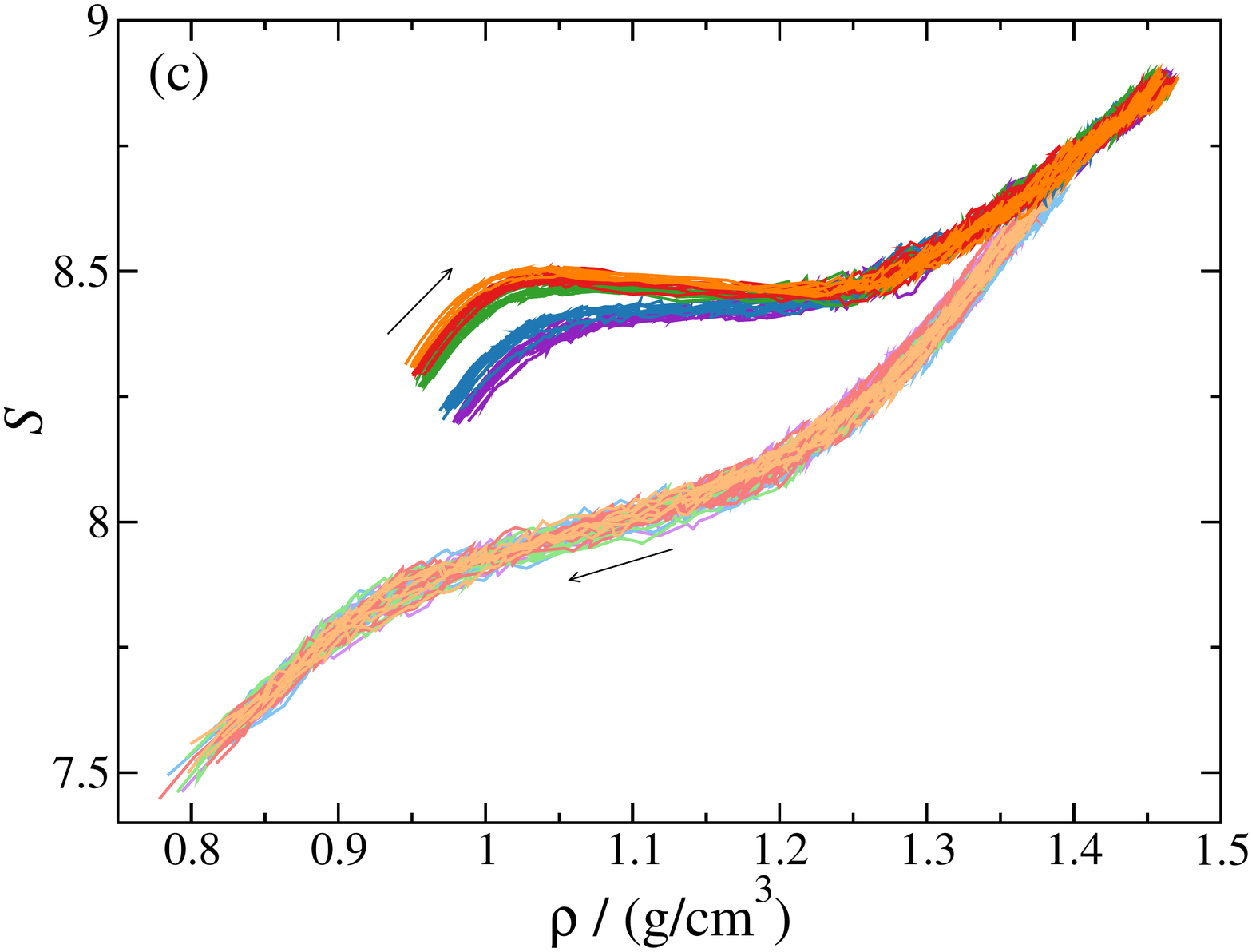}
\caption{
(a) Pressure, (b) energy, and (c) shape function of the IS sampled by the system during the compression/decompression cycles shown in Fig.~\ref{fig:rho-qdep}.
During compression, an anomalous concavity region develops in  $e_\text{IS}(\rho)$ with decreasing $q_\text{c}$, and anomalous van der Waals-like loops become observable in $P_\text{IS}(\rho)$ and $\mathcal{S}(\rho)$.
 All PEL properties change monotonically with density during decompression.
}
\label{fig:pel-qdep}
\end{center}
\end{figure}

We summarize these results in Fig.~\ref{fig:trans-qdep}, where the pressure of the LDA$\rightarrow$HDA transformation is shown as function of $1/q_\text{c}$.
For comparison, we also include available experimental data.
It follows from Fig.~\ref{fig:trans-qdep} that as $1/q_\text{c}$ increases and $q_\text{c}$ approaches the experimental rate, the LDA$\rightarrow$HDA transformation pressure seems to reach an asymptotic  value of $\approx950$~MPa.
Although this pressure is larger than the corresponding experimental pressure of $\approx550$~MPa, one should note that the compression rates in experiments and MD simulation are very different (the role of $q_\text{P}$ on our systems is discussed in the next section).
We note that, the pressure associated to the  HDA$\rightarrow$LDA transformation is not shown in Fig.~\ref{fig:trans-qdep}, because, at the present $T$ and $q_\text{P}$, this transformation is very smooth.

\begin{figure}[t!]
\begin{center}
\includegraphics[width=\columnwidth]{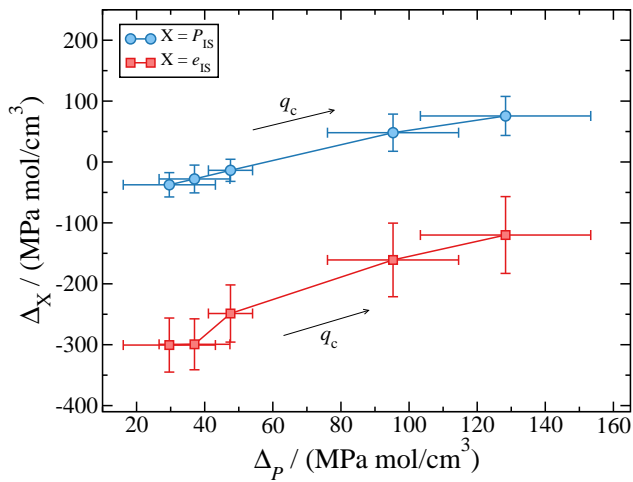}
\caption{
Relationship between the sharpness of the LDA$\rightarrow$HDA transformations shown in Fig.~\ref{fig:rho-qdep} and the corresponding anomalous character of the PEL properties (Fig.~\ref{fig:pel-qdep}). 
$\Delta_P$ and  $\Delta_{P_\text{IS}}$ are the slopes of $P(\rho)$ and $P_\text{IS}(\rho)$ during the LDA$\rightarrow$HDA transformations (see Eqs.~\ref{eq:delta} and \ref{eq:delta-pis}).
$\Delta_{e_\text{IS}}$ quantifies the concavity in $e_\text{IS}$ during the transformation (see Eq.~\ref{eq:delta-eis}).
As $\Delta_P\rightarrow 0$ and the transformation becomes sharper (reminiscent of a first-order phase transition), $\Delta_{P_\text{IS}}$ decreases and becomes negative, i.e., $P_\text{IS}(\rho)$ develops an anomalous van der Waals-like loop (Fig.~\ref{fig:pel-qdep}).
$\Delta_{e_\text{IS}}$ is negative, i.e., $e_\text{IS}(\rho)$ is anomalously concave, and becomes more negative as the sharpness of the LDA$\rightarrow$HDA transformation becomes more pronounced.
}
\label{fig:delta-qdep}
\end{center}
\end{figure}

The PEL properties sampled by the system during the compression/decompression cycles in Fig.~\ref{fig:rho-qdep} are shown in Fig.~\ref{fig:pel-qdep}.
During compression, $P_\text{IS}(\rho)$ increases monotonically for LDA forms prepared with fast cooling rates, $q_\text{c}\geq30$~K/ns [Fig.~\ref{fig:pel-qdep}(a)]. 
However, as $q_\text{c}$ decreases, a clear van der Waals-like loop develops in $P_\text{IS}(\rho)$ during the LDA$\rightarrow$HDA transformation.
In particular, this van der Waals-like loop becomes more pronounced as the slope of $\rho(P)$ in Fig.~\ref{fig:rho-qdep} becomes sharper.  
This is clearly shown in Fig.~\ref{fig:delta-qdep} where the $\Delta_{P_\text{IS}}$ is plotted as function of $\Delta_P$.    
Interestingly, during the decompression of HDA, $P_\text{IS}(\rho)$ decreases monotonically (i.e., it exhibits no van der Waals-like loop), which is also consistent with the smooth behavior of $\rho(P)$ shown in Fig.~\ref{fig:rho-qdep} along the decompression paths.
We also note that the behavior of $P_\text{IS}(\rho)$ during decompression of HDA is independent of $q_\text{c}$.
This, again, is consistent with the assessment  that the system looses memory once it reaches the HDA state.

The behavior of $e_\text{IS}(\rho)$ [Fig.~\ref{fig:pel-qdep}(b)], is fully consistent with the evolution of $P_\text{IS}(\rho)$ during the compression/decompression cycles.
Specifically, during compression $e_\text{IS}(\rho)$ exhibits a concavity region during the LDA$\rightarrow$HDA transformation that becomes more pronounced as $q_\text{c}$ decreases.
In other words, as the van der Waals-like loop in $P_\text{IS}(\rho)$ becomes more pronounced [see Fig.~\ref{fig:pel-qdep}(a)], and the LDA$\rightarrow$HDA transformation becomes sharper (see Fig.~\ref{fig:rho-qdep}), $e_\text{IS}(\rho)$ becomes increasingly a more concave function of $\rho$.
This is also visible in Fig.~\ref{fig:delta-qdep} where the $\Delta_{e_\text{IS}}$ is plotted as function of $\Delta_P$.
Instead, during decompression of HDA, $e_\text{IS}(\rho)$ shows no concavity which is consistent with the lack of a van der Waals-like loop in $P_\text{IS}(\rho)$  [see Fig.~\ref{fig:pel-qdep}(a)] and the smooth behavior of $\rho(P)$ (see Fig.~\ref{fig:rho-qdep}) along the corresponding path.

A subtle point follows from Fig.~\ref{fig:pel-qdep}(b).
Specifically, the $e_\text{IS}(\rho)$ corresponding to the starting LDA forms at $\rho\approx0.95$--$0.97$~g/cm$^3$ become more negative with decreasing $q_\text{c}$.
In other words, as the cooling rate decreases during the  preparation of LDA, the system is able to relax for longer times during the cooling process, and the final LDA state is able to reach deeper regions of the PEL.
Accordingly, our results imply that in order to observe a first-order-like phase transition during the compression of LDA, one should start with LDA forms located deep within the LDA region of the PEL (cf. Ref.~\onlinecite{giovambattista17-jcp} for the case of glassy ST2 water).
From a microscopic point of view, the LDA forms with lower $e_\text{IS}(\rho)$ are characterized by higher tetrahedral order as is obvious from the structure factor shown in Fig.~\ref{fig:sq-qdep} (cf. also Refs.~\onlinecite{engstler17-jcp,martelli18-prm}).
 This high degree of tetrahedrally makes the hydrogen-bond network of water stronger and more resistant to collapse under pressure during the LDA$\rightarrow$HDA transformation.
 Thus, it is the high tetrahedral order in LDA what is required to observe a sharp first-order like transition.

Regarding the curvature of the IS sampled by the system [Fig.~\ref{fig:pel-qdep}(c)], we find that the shape function $\mathcal{S}(\rho)$ follows closely the behavior of $P_\text{IS}(\rho)$ during the compression/decompression cycles.
For example, when $P_\text{IS}(\rho)$ shows a van der Waals-like loop, $\mathcal{S}(\rho)$ does it as well.
When $P_\text{IS}(\rho)$ is a monotonic function of $\rho$,  $\mathcal{S}(\rho)$ is also a monotonic function.
This parallel behavior of $P_\text{IS}(\rho)$ and $\mathcal{S}(\rho)$ was also noted in Sec.~\ref{sec:tdep}.

\subsubsection{Compression Rate Dependence}
\label{sec:pidep}

\begin{figure}[t!]
\begin{center}
\includegraphics[width=\columnwidth]{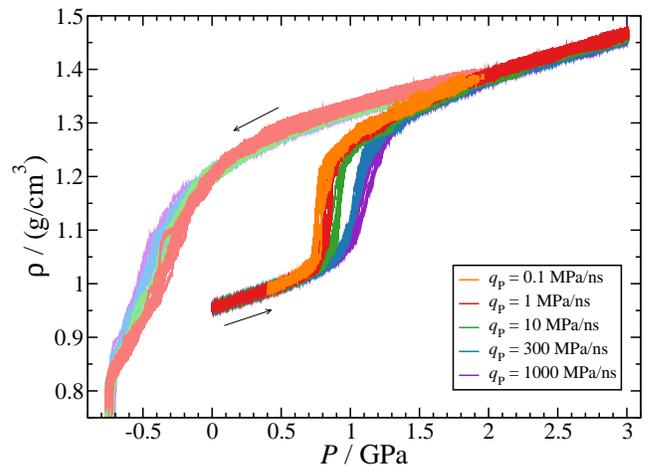}
\caption{
Density as function of pressure during the  compression of LDA and subsequent decompression of HDA (from $P=2$~GPa, lighter colors) at $T=80$~K.
LDA is prepared by isobaric cooling at $P=0.1$~MPa using using a cooling rate  $q_\text{c}=0.1$~K/ns.
Different compression/decompression  rates $q_\text{P}=0.1$--$1000$~MPa/ns are used.
}
\label{fig:rho-pidep}
\end{center}
\end{figure}

Next, we study the effects of varying $q_\text{P}$ on the LDA$\rightleftarrows$HDA transformation. 
All LDA samples considered in this section are prepared by isobaric cooling with $q_\text{c}=0.1$~K/ns and the compression/decompression runs are performed at $T=80$~K.
The behavior of $\rho(P)$ during these runs ($q_\text{p}=0.1$--1000~MPa/ns) is shown in Fig.~\ref{fig:rho-pidep}.
It is visible that decreasing $q_\text{P}$ makes the LDA$\rightarrow$HDA transformation much sharper and it shifts it to lower $P$.
This is a reasonable behavior since the system has more time to relax during the transformation as $q_\text{P}$ is reduced.
Similarly, during the decompression process, the HDA$\rightarrow$LDA transformation also becomes slightly more evident
and shifts to higher (less negative) $P$ with decreasing values of $q_\text{P}$.

\begin{figure}[tb]
\begin{center}
\includegraphics[width=\columnwidth]{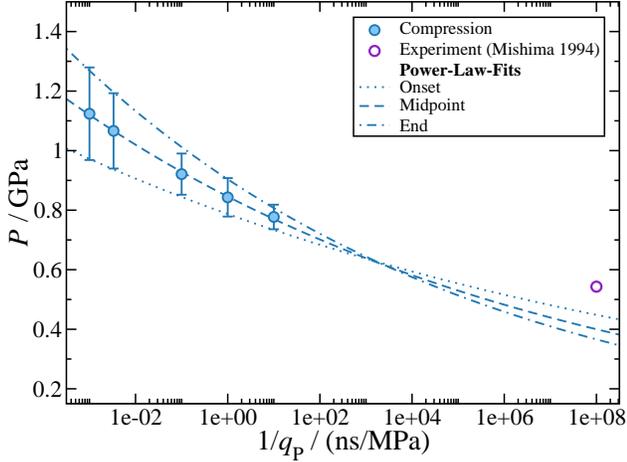}
\caption{
Pressure as function of the inverse of the compression rate corresponding to the LDA$\rightarrow$HDA transformations shown in Fig.~\ref{fig:rho-pidep}.
The circles and associated `error bars' represent, respectively, the midpoint and the width of the corresponding transformation.
The purple circle indicates the experimental LDA$\rightarrow$HDA transformation pressure obtained in the experiments of Ref.~\onlinecite{mishima94-jcp}.
Our data extrapolate fairly well to the experimental point using a power-law fit.
}
\label{fig:trans-pidep}
\end{center}
\end{figure}

The results from Fig.~\ref{fig:rho-pidep} are summarized in Fig.~\ref{fig:trans-pidep} where we also include available experimental data.
It is again visible that the LDA$\rightarrow$HDA transformation pressure decreases and the density steps become steeper as $q_\text{P}$ decreases.
In particular, the MD data extrapolates reasonably well to the experimental data from Ref.~\onlinecite{mishima94-jcp}.
Unfortunately there is no systematic experimental study of the $q_\text{P}$ dependence of the LDA$\rightarrow$HDA transformation at 80~K.
At 125~K, however, experiments found no significant change in the transformation pressure as the rate was increased from $0.1$~MPa/s to $100$~MPa/s~\cite{loerting06-prl}.
During decompression we find no HDA$\rightarrow$LDA transition at positive pressures for all rates studied (please note that the smallest $q_\text{P}$ used for the decompressions is 1~MPa/ns), a finding consistent with experiments~\cite{mishima94-jcp}.
Interestingly, the MD simulation data in Fig.~\ref{fig:rho-pidep} shows that the hysteresis in $~\rho(P)$ during the LDA$\rightleftarrows$HDA transformation becomes smaller as $q_\text{P}$ decreases.
It seems, however, that changes in $q_\text{P}$ affect the LDA$\rightarrow$HDA more than the HDA$\rightarrow$LDA transformation.

 \begin{figure}[t!]
\begin{center}
\includegraphics[width=0.96\columnwidth]{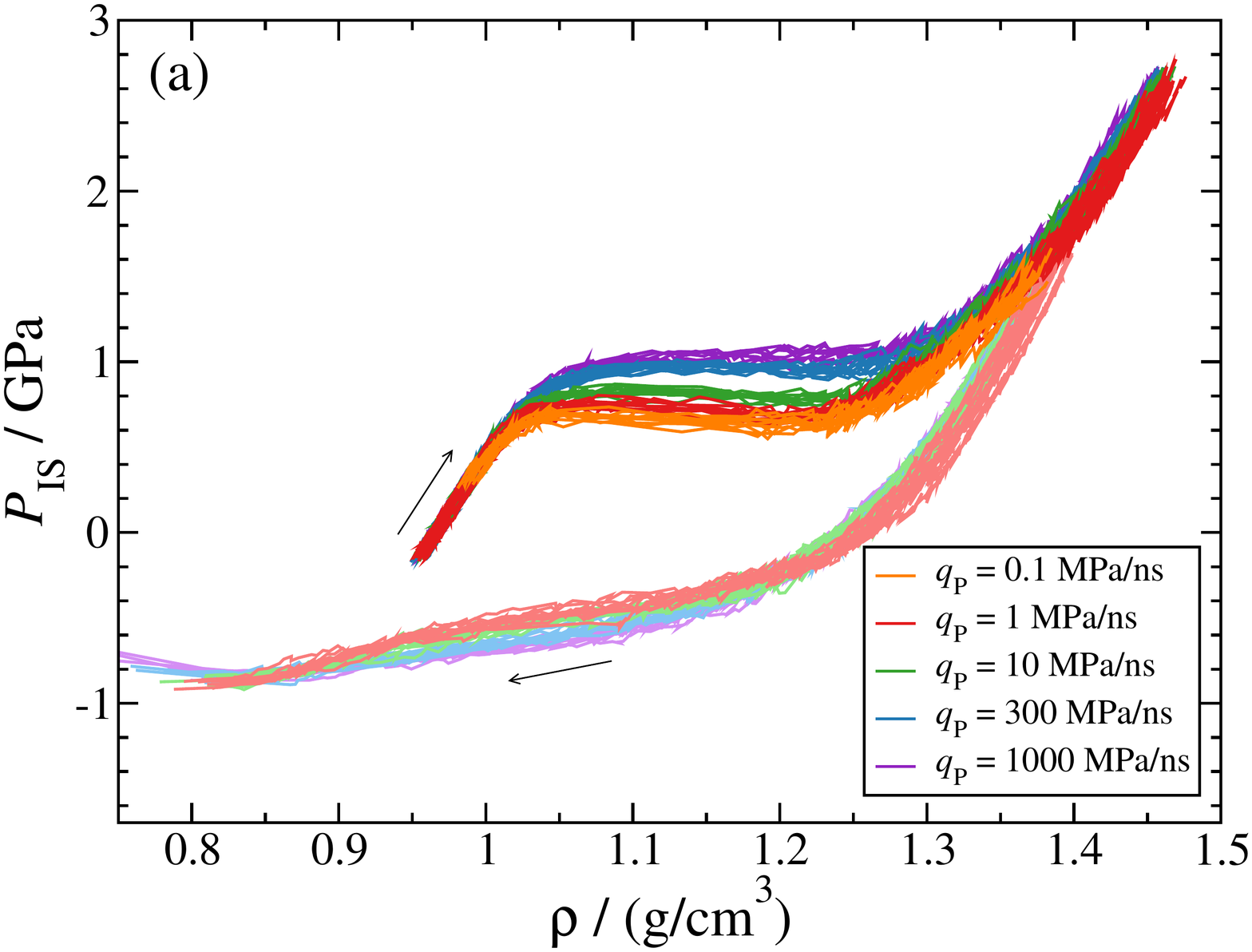}
\includegraphics[width=0.96\columnwidth]{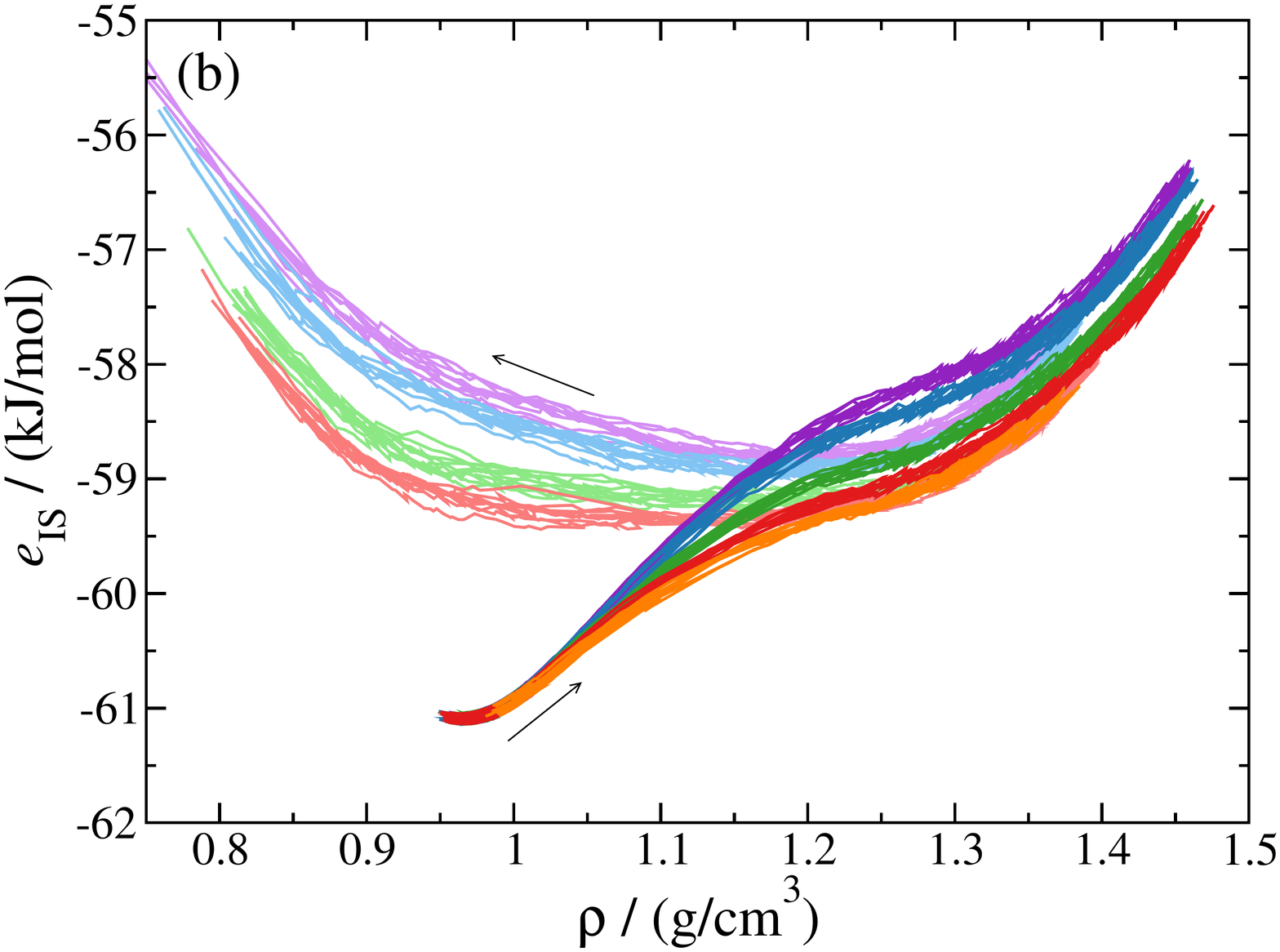}
\includegraphics[width=0.96\columnwidth]{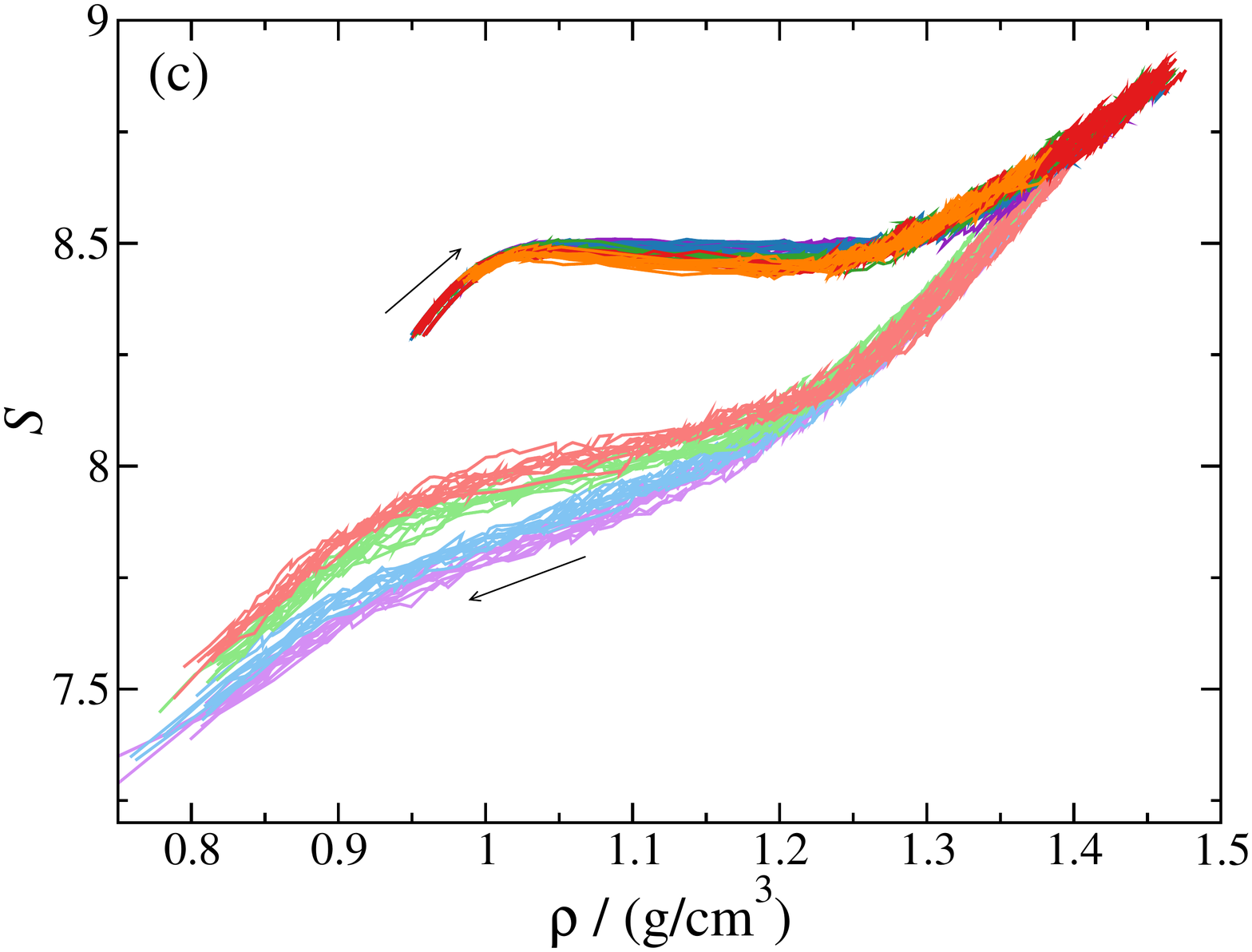}
\caption{
(a) Pressure, (b) energy, and (c) shape function of the IS sampled by the system during the compression/decompression cycles shown in Fig.~\ref{fig:rho-pidep}.
During compression, an anomalous concavity region is present in $e_\text{IS}(\rho)$
and anomalous van der Waals-like loops become observable in $P_\text{IS}(\rho)$ and $\mathcal{S}(\rho)$ as $q_\text{P}$ is decreassed.
 All PEL properties change monotonically with density during decompression.
}
\label{fig:pel-pidep}
\end{center}
\end{figure}
 
\begin{figure}[t!]
\begin{center}
\includegraphics[width=\columnwidth]{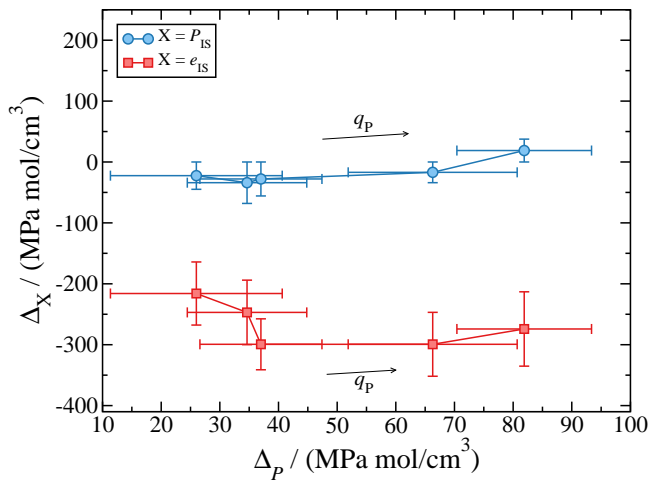}
\caption{
Relationship between the sharpness of the LDA$\rightarrow$HDA transformations shown in Fig.~\ref{fig:rho-pidep} and the corresponding anomalous character of the PEL properties (Fig.~\ref{fig:pel-pidep}). 
$\Delta_P$ and  $\Delta_{P_\text{IS}}$ are the slopes of $P(\rho)$ and $P_\text{IS}(\rho)$ during the LDA$\rightarrow$HDA transformations (see Eqs.~\ref{eq:delta} and \ref{eq:delta-pis}).
$\Delta_{e_\text{IS}}$ quantifies the concavity in $e_\text{IS}$ during the transformation (see Eq.~\ref{eq:delta-eis}).
As $\Delta_P\rightarrow 0$ and the transformation becomes sharper (reminiscent of a first-order phase transition), $\Delta_{P_\text{IS}}$ decreases and becomes negative, i.e., $P(\rho)$ develops an anomalous van der Waals-like loop (Fig.~\ref{fig:pel-pidep}).
$\Delta_{e_\text{IS}}$ is negative at all rates studied, i.e., $e_\text{IS}(\rho)$ is anomalously concave.
}
\label{fig:delta-pidep}
\end{center}
\end{figure}

The behavior of $P_\text{IS}(\rho)$, $e_\text{IS}(\rho)$ and $\mathcal{S}(\rho)$ during the compression/decompression cycles in Fig.~\ref{fig:rho-pidep} are shown in Fig.~\ref{fig:pel-pidep}.
During compression, a van der Waals-like loop develops in $P_\text{IS}(\rho)$ [see Fig.~\ref{fig:pel-pidep}(a)] as well as in $\mathcal{S}_\text{IS}(\rho)$ [see Fig.\ref{fig:pel-pidep}(c)] for decreasing values of   $q_\text{p}$.
In particular, we note that the anomalous behavior in $P_\text{IS}(\rho)$ becomes more pronounced as the LDA$\rightarrow$HDA transformation becomes sharper.
In addition, $e_\text{IS}(\rho)$ shows a concavity region during the LDA$\rightarrow$HDA transformation [see Fig.\ref{fig:pel-pidep}(b)].
These findings are also evident in Fig.~\ref{fig:delta-pidep}, where we plot $\Delta_{P_\text{IS}}$ and $\Delta_{e_\text{IS}}$ as a function of $\Delta_P$.
Here we note a slight increase of $\Delta_{e_\text{IS}}$ at low $\Delta_P$ although it is unclear if this increase is indeed significant given the variance of the data.

During decompression, there is no anomalous behavior in any of the PEL properties studied.
 $P_\text{IS}(\rho)$ and $\mathcal{S}(\rho)$ decay monotonically during decompression and $e_\text{IS}(\rho)$ has positive curvature at all densities.
 This is consistent with the smooth decrease in $\rho(P)$ during the decompression of HDA (see Fig.~\ref{fig:rho-pidep}). 
We note, however, that as the HDA$\rightarrow$LDA transformation becomes sharper with decreasing $q_\text{P}$, the slope of  $P_\text{IS}(\rho)$ at the transformation seems to approach zero.
 One may expect that as $q_\text{P}$ decreases below 1~MPa/ns,  $P_\text{IS}(\rho)$ may also exhibit a van der Waals-like loop during the HDA$\rightarrow$LDA transformation. 
 A consistent trend follows from Fig.~\ref{fig:pel-pidep}(b) for the case of $e_\text{IS}(\rho)$. During the decompression path, the curvature at $\rho\approx1.1$~g/cm$^3$ decreases with decreasing $q_\text{P}$ and, for $1$~MP/ns, the curvature of $e_\text{IS}(\rho)$ is practically zero.
 Hence we expect that, $e_\text{IS}(\rho)$ should develop a concavity during the HDA$\rightarrow$LDA transformation for $q_\text{P} < 1$~MPa/ns. 
We note that the sharpness of the HDA$\rightarrow$LDA transformation during the decompression path is sensitive to $q_\text{P}$, but not $q_\text{c}$.
 Not surprisingly, only reducing $q_\text{P}$ allows the system to increasingly relax during the HDA$\rightarrow$LDA transformation.
 This is indicated by the IS energies explored by the system during decompression in Fig.~\ref{fig:pel-pidep}(b).
 As $q_\text{P}$ decreases, the $e_\text{IS}(\rho)$ of the system at $\approx 0.95$~g/cm$^3$ decreases meaning that the system accesses LDA configurations that are located deeper in the LDA state and closer to the starting LDA.
 Instead, as shown in Fig.~\ref{fig:pel-qdep}(b), the system reaches the same $e_\text{IS}$  at $\rho\approx 0.9$~g/cm$^3$ during the  decompression of HDA for all cooling rates $q_\text{c}$ considered.

\section{Conclusions}
\label{sec:concl}

In this report, we explored the PEL of TIP4P/2005 water during the pressure induced LDA$\rightarrow$HDA and HDA$\rightarrow$LDA transformations.
The initial LDA form for the compression runs was produced by quenching the liquid at $P=0.1$~MPa with cooling rates as low as $0.01$~K/ns.
This cooling rate coincides with rates reached in hyperquenching experiments~\cite{bruggeller80-nature, dubochet81-jmic, mayer82-nature, mayer85-jap,kohl00-pccp}.
Reducing the cooing rate from 100~K/ns to 0.01~K/ns allows the system to access deeper and deeper regions of the PEL.
From a microscopic point of view, the deeper the system is within the LDA region of the PEL, the more tetrahedral it is.
At our slowest cooling rate, the density of LDA is slightly above the density of TIP4P/2005 ice I$_\text{h}$.

During the compression-induced LDA$\rightarrow$HDA transformation, a pronounced density increase occurs.
The sharpness of this density increase was found to be strongly dependent on the LDA preparation process (i.e., cooling rate $q_\text{c}$) as well as the compression rate $q_\text{P}$ and temperature $T$.
At $T<80$~K, the  LDA$\rightarrow$HDA transformation is rather smooth, due to the slow kinetics of the transformation and relatively fast  compression rates employed.
However, as $T\rightarrow T_\text{c}$, the LDA$\rightarrow$HDA transformation becomes sharp, reminiscent of a first-order phase transition, as observed in experiments.
By studying the compression-induced LDA$\rightarrow$HDA transformation at fixed temperature and compression rate 
($T=80$~K, $q_\text{P}$=10~MPa/ns), we find that reducing $q_\text{c}$ leads to a sharper transformation between LDA and HDA. In other words, as the tetrahedrality of LDA increases, the LDA$\rightarrow$HDA transformation becomes more reminiscent of a first-order phase transition.
Similarly, by studying the compression-induced LDA$\rightarrow$HDA transformation at fixed temperature and cooling rate 
($T=80$~K, $q_\text{c}$=0.1~K/ns), we find that reducing $q_\text{P}$ leads to a sharper transition between LDA and HDA.
Remarkably, our LDA$\rightarrow$HDA transformation pressures, obtained at different compression rates, extrapolate fairly well to the experimental transformation pressure.

During the decompression of HDA, we can only observe an HDA$\rightarrow$LDA transformation at $T=160$~K, close to the estimated LLCP.
At lower $T$, the transformation is rather smooth.
However, we also find that as the decompression rate $q_\text{P}$ decreases, an HDA$\rightarrow$LDA transformation becomes more apparent in the behavior of $\rho(P)$.
Interestingly, the HDA$\rightarrow$LDA transformation is insensitive to the cooling rate employed in the  preparation of LDA.   Consistent with the case of ST2 water~\cite{giovambattista16-jcp,giovambattista17-jcp}, our results indicate that once HDA forms (at $P=2$~GPa), the system seems to completely loose memory of its history.

In agreement with previous studies~\cite{giovambattista16-jcp,giovambattista17-jcp,sun18-prl}, we find that at those conditions ($T$, $q_\text{c}$, $q_\text{P}$) where
the LDA$\rightarrow$HDA transformation is reminiscent of a first-order-like phase transition, the PEL properties sampled by the system during the transformation become anomalous.
Specifically, during the LDA$\rightarrow$HDA transformation, (i) $e_\text{IS}(\rho)$ becomes a concave function of $\rho$ and (ii) a van der Waals-like loop develops in $P_\text{IS}(\rho)$.
In addition, and in agreement with results obtained for ST2 water~\cite{giovambattista16-jcp,giovambattista17-jcp}, we also find that ${\cal S}(\rho)$ is anomalous, exhibiting also a van der Waals-like loop.
These features are very weak or absent during  smooth HDA$\rightarrow$LDA transformations. 

Our studies at $T=80$~K and different rates ($q_\text{c}$, $q_\text{P}$), show that the anomalous 
van der Waals-like loop in $P_\text{IS}(\rho)$ becomes more pronounced as the LDA$\rightarrow$HDA transformation becomes sharper, i.e.,
 more reminiscent of a first-order phase transition.
The case of $e_\text{IS}(\rho)$ is less clear but our MD simulations show that $e_\text{IS}(\rho)$  remains a concave function of $\rho$ at all conditions studied. 
We argue that the PEL anomalies (i) and (ii) are necessary but not sufficient conditions for a system to exhibit a fist-order-like phase transition between LDA and HDA forms.
Indeed, as discussed in Sec.~\ref{sec:pel-liq}, for the case of a supercooled liquid (e.g., close to the glass transition temperature), these anomalies of the PEL may originate a first-order phase transition between two liquid states.  

Previously similar PEL studies were conducted for the SPC/E and ST2 models of water~\cite{giovambattista03-prl,giovambattista16-jcp, giovambattista17-jcp}.
The main difference between  SPC/E and ST2 water is that an LLCP is accessible in (metastable) equilibrium simulations of ST2 water~\cite{poole92-nature, poole05-jpcm,cuthbertson11-prl, liu12-jcp, smallenburg2015tuning, palmer14-nature, palmer18-jcp, palmer18-crev}, while an LLCP is not accessible in SPC/E water~\cite{scala00-nature,scala00-pre,sciortino03-prl}.
Consistent with this difference the PEL of SPC/E water shows smooth changes during the LDA$\rightleftarrows$HDA transformations including a very weak, concavity in $e_\text{IS}$ ~\cite{giovambattista03-prl}.
In ST2 on the other hand,  van der Waals-like loops in $P_\text{IS}(\rho)$ and $\mathcal{S}(\rho)$ as well as a concavity in $e_\text{IS}(\rho)$ are present~\cite{giovambattista16-jcp, giovambattista17-jcp}.
Even a maximum in $e_\text{IS}(\rho)$ was reported, consistent with the presence of two megabasins~\cite{giovambattista16-jcp, giovambattista17-jcp}.
That is, the signs expected for a first-order-like phase transition in the PEL are significantly weaker in SPC/E water than in ST2 water.
This suggests that the glass phenomenology observed in SPC/E water can be thought of as ``supercritical", analogous to a liquid$\rightleftarrows$gas transformation at $T>T_\text{c}$. 
For ST2 water, the glass phenomenology resembles more a ``subcritical" first-order phase transition.
In our study of  TIP4P/2005 water we find PEL features similar to ST2 water, including van der Waals-like loops in $P_\text{IS}(\rho)$ and $\mathcal{S}(\rho)$ as well as a concavity in $e_\text{IS}(\rho)$.
However, we find no maximum in $e_\text{IS}(\rho)$ for TIP4P/2005 water.
That is, in the case of TIP4P/2005 water, the two distinct regions of the PEL associated to LDA and HDA are not necessarily two different megabasins of the PEL.
Similar conclusions were found in Ref.~\onlinecite{sun18-prl} for a water-like model system.
We stress that our study of TIP4P/2005 water is consistent with studies indicating an LLCP in this model~\cite{abascal10-jcp, wikfeldt2011spatially, sumi13-rscadv, yagasaki14-pre, russo14-natcomm, singh16-jcp, biddle17-jcp, handle18-jcp}.

We conclude by noticing that the amorphous ices sample regions of the PEL that are different from the ones sampled by the equilibrium liquid (see supplementary material).
This has profound implications in the relationship between the liquid and glass state.
Specifically, this implies that the PEL regions sampled by the equilibrium liquid and the glass state differ.
It follows that it may not be possible to predict quantitatively the behavior of the glass state based only on properties of the equilibrium liquid.
In this work, we found that, indeed, there are no direct correlations between the LDL$\rightleftarrows$HDL spinodal lines of TIP4P/2005 liquid water obtained from the PEL-EOS of Ref.~\onlinecite{handle18-jcp} and the corresponding  transformation pressures between LDA and HDA.
However, in the LLCP scenario, the LDA$\rightleftarrows$HDA transformation lines are extensions of the LDL$\rightleftarrows$HDL spinodal lines into the glass domain.

\section*{Supplementary Material}

In the supplementary material we compare the PEL regions sampled during the pressure induced LDA$\rightleftarrows$HDA transformations  to the PEL regions sampled by the equilibrium liquid.
We show that the regions of the PEL sampled by the equilibrium liquid and the amorphous ices are, in general, different.

\begin{acknowledgments}
PHH thanks the Austrian Science Fund FWF (Erwin Schr\"{o}dinger Fellowship J3811 N34) and the University of Innsbruck (NWF-Project 282396) for support.
FS acknowledges support from MIUR-PRIN grant 2017Z55KCW.
The computational results presented have been achieved using the HPC infrastructure LEO of the University of Innsbruck and the HPCC of CUNY.
The CUNY HPCC is operated by the College of Staten Island and funded, in part, by grants from the City of New York, State of New York, CUNY Research Foundation, and National Science Foundation Grants CNS-0958379, CNS-0855217 and ACI 1126113.

\end{acknowledgments}

\bibliography{tip4p2005-polyam.bib}

\end{document}


\title{Supplementary Material: ``Glass Polymorphism in TIP4P/2005 Water:  A Description Based on the Potential Energy Landscape Formalism"\\
}


\author{Philip H. Handle}
\affiliation{%
 Institute of Physical Chemistry,
 University of Innsbruck,
 Innrain 52c, A-6020 Innsbruck, Austria
}
\author{Francesco Sciortino}%
\affiliation{%
 Department of Physics,
 Sapienza -- University of Rome,
 Piazzale Aldo Moro 5, I-00185 Roma, Italy
}
\author{Nicolas Giovambattista}
\affiliation{
 Department of Physics,
 Brooklyn College of the City University of New York,
 New York, New York 10016, USA}
\affiliation{Ph.D. Programs in Chemistry and Physics,
 The Graduate Center of the City University of New York,
 New York, New York 10016, USA
}

\date{\today}

\maketitle

In this supplementary material we compare the PEL regions sampled by the system in the (stable/metastable) equilibrium liquid state and during the pressure-induced LDA$\rightleftarrows$HDA transformations.
The data for the equilibrium liquid stem from Ref.~\onlinecite{handle18-jcp}.
We follow the structure of the main document and discuss, separately, the influence of temperature $T$, cooling rate $q_\text{c}$ and compression rate $q_\text{P}$. 

\section{Temperature Dependence}

\begin{figure}[b!]
\begin{center}
\includegraphics[width=0.49\textwidth]{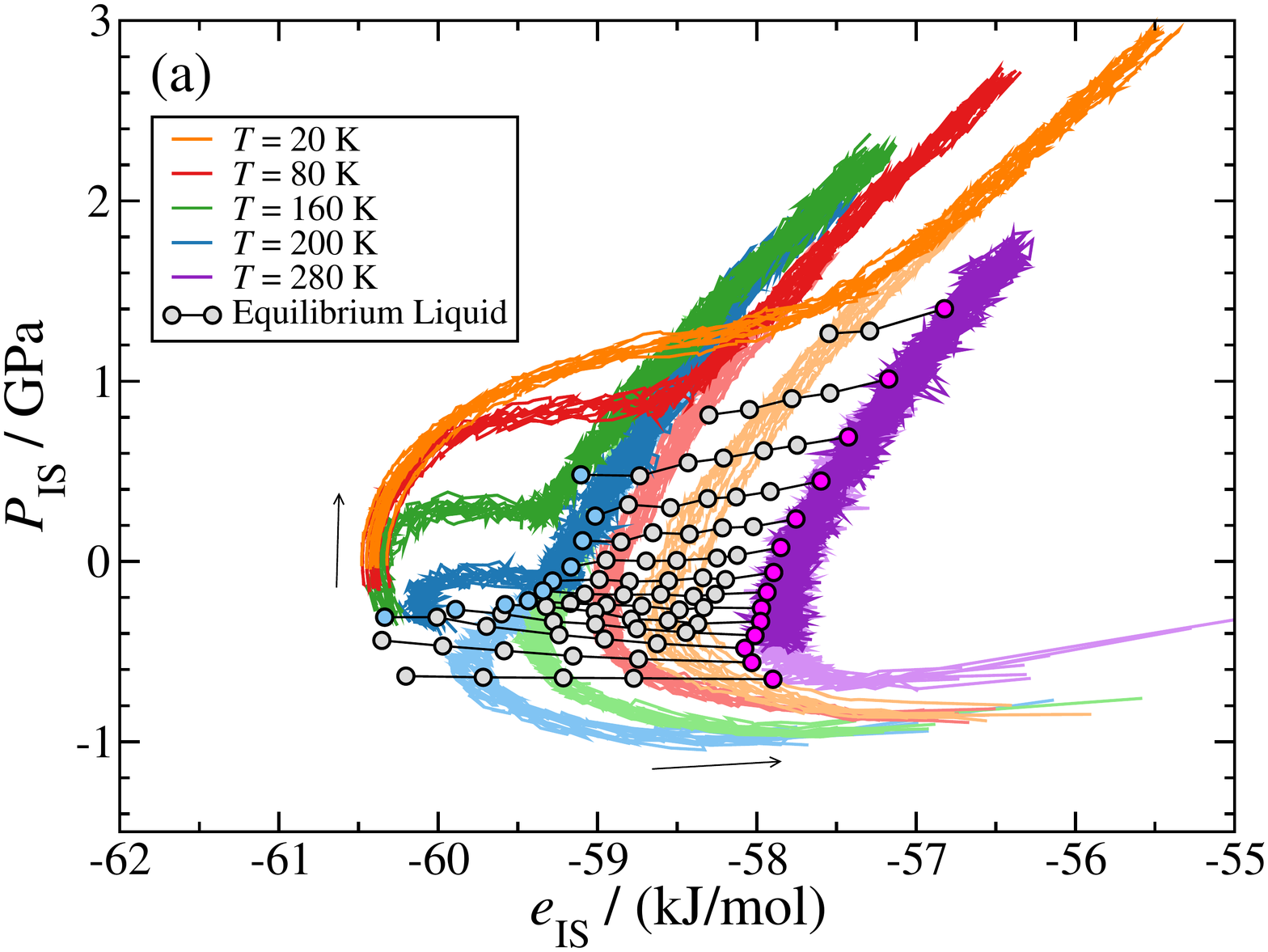}
\includegraphics[width=0.49\textwidth]{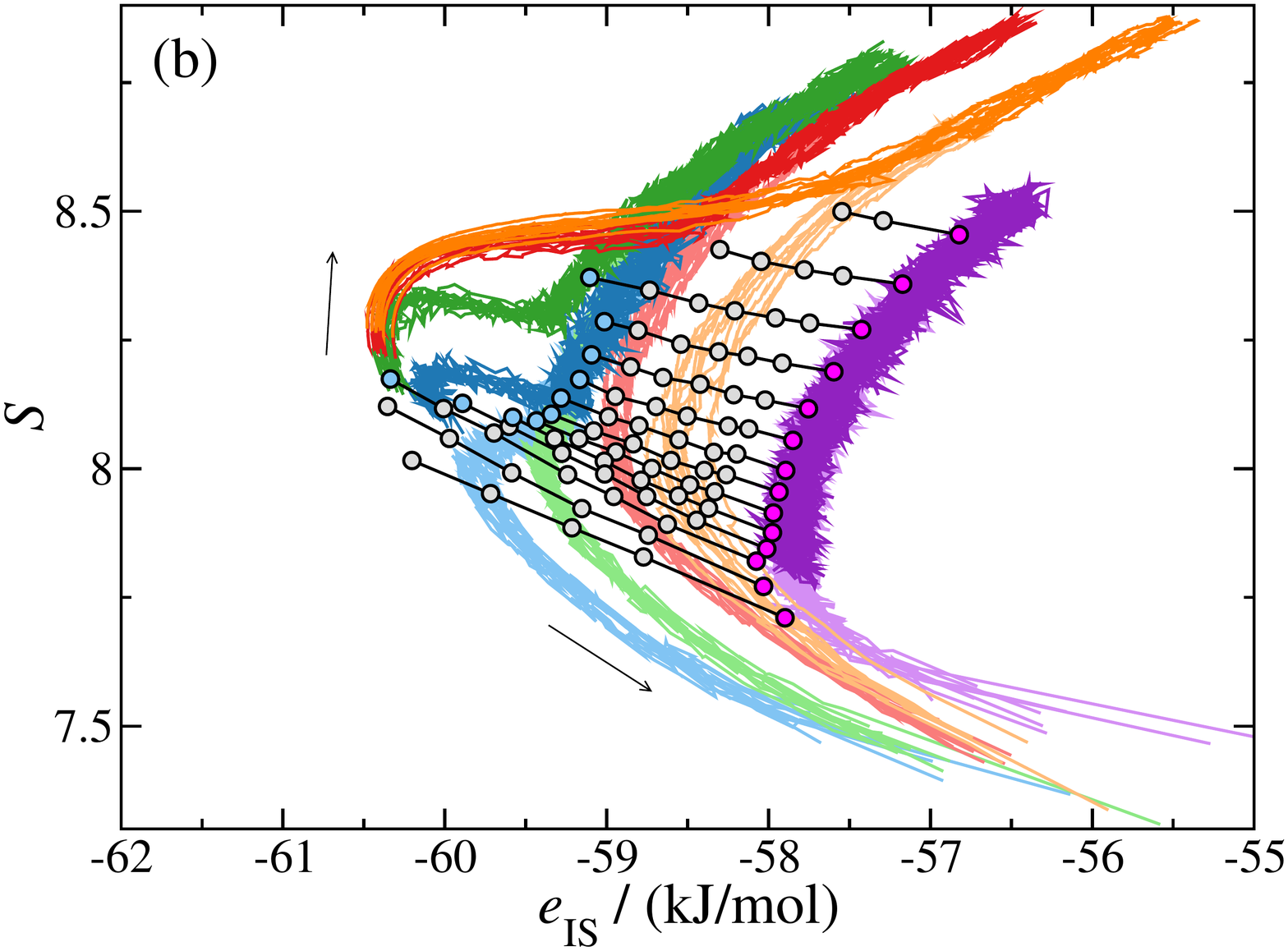}
\caption{
(a) IS Pressure, $P_\text{IS}$, and (b) shape function, $\mathcal S$, as a function of the IS energy sampled by the system during the pressure-induced LDA$\rightleftarrows$HDA transformation (solid lines).
Sharp and light colors correspond, respectively,  to the LDA$\rightarrow$HDA and HDA$\rightarrow$LDA transformations.
Data is taken from Fig.~6 of the main manuscript.
LDA is prepared by isobaric cooling at $P=0.1$~MPa using a cooling rate $q_\text{c}=30$~K/ns.
The compression/decompression rate is $q_\text{P}=300$~MPa/ns.
The circles represent the $P_\text{IS}(e_\text{IS})$ and $\mathcal{S}(e_\text{IS})$ for the stable/metastable equilibrium liquid from Ref.~\onlinecite{handle18-jcp} with densities in the range $0.90$--$1.42$~g/cm$^3$  (bottom to top lines, in steps of 0.04~g/cm$^3$) and temperatures from $T\geq200$~K (blues circles) to $T=270$K (magenta circles).
At low temperatures,  the PEL regions sampled by the equilibrium liquid and the amorphous ices are different.
}
\label{fig:delta-tdep}
\end{center}
\end{figure}

Fig.~\ref{fig:delta-tdep}(a) shows the pressure of the inherent structures (IS) sampled by the system during the LDA$\rightleftarrows$HDA transformation as a function of the corresponding IS energy, $P_\text{IS}(e_\text{IS})$.
For comparison, also included is the $P_\text{IS}(e_\text{IS})$ of the liquid in (stable and metastable) equilibrium at $200\leq T \leq270$~K and $0.90\leq\rho\leq1.42$~g/cm$^3$ taken from Ref.~\onlinecite{handle18-jcp} (circles).
Similarly, Fig.~\ref{fig:delta-tdep}(b) shows the shape function as function of the IS energy, $\mathcal{S}(e_\text{IS})$, sampled by the system in liquid state and during the LDA$\rightleftarrows$HDA transformation.
It follows from Fig.~\ref{fig:delta-tdep} that the system samples different regions of the PEL during compression (sharp color lines) and decompression (light color lines).

At high compression/decompression temperatures, within the liquid state, the system samples the same IS accessed by the liquid in equilibrium.
For example, the values of $P_\text{IS}(e_\text{IS})$ and $\mathcal{S}(e_\text{IS})$ for the equilibrium liquid at $T=270$~K (magenta circles) practically coincide with the corresponding values obtained here during compression/decompression at $T=280$~ K (violet lines).
However, at low temperatures, the amorphous ices at $T=20$--80~K (orange and red lines) can access IS with $e_\text{IS}<-60$~kJ/mol and $P_\text{IS}>0.5$~GPa while IS sampled by the liquid with $e_\text{IS}<-60$~kJ/mol are characterized by $P_\text{IS}<0$. 
Similar conclusions follow from Fig.~\ref{fig:delta-tdep}(b) regarding $\mathcal{S}(e_\text{IS})$.  

\section{Cooling Rate Dependence}
\label{sec:qdep}

\begin{figure}[h!]
\begin{center}
\includegraphics[width=0.49\textwidth]{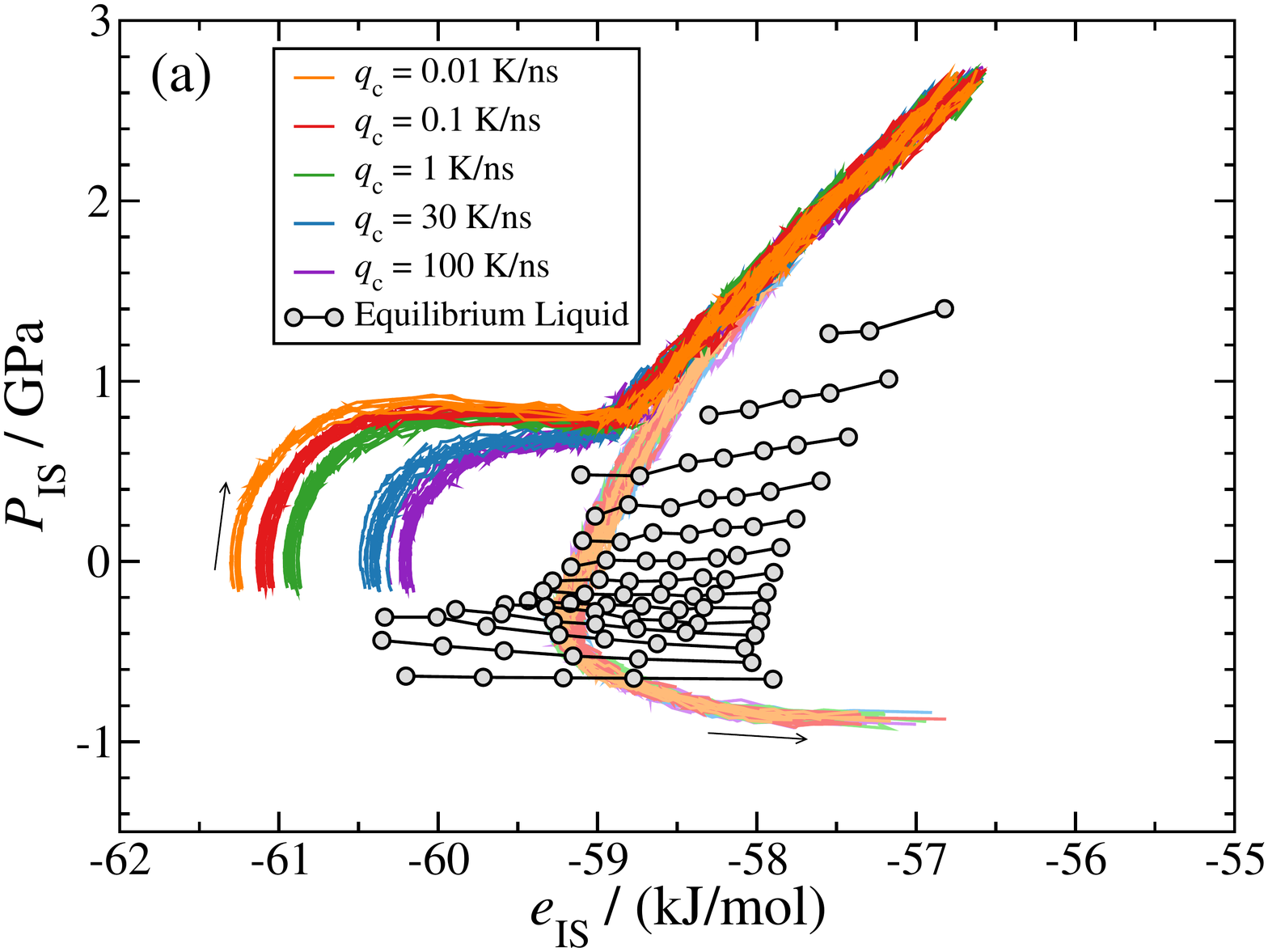}
\includegraphics[width=0.49\textwidth]{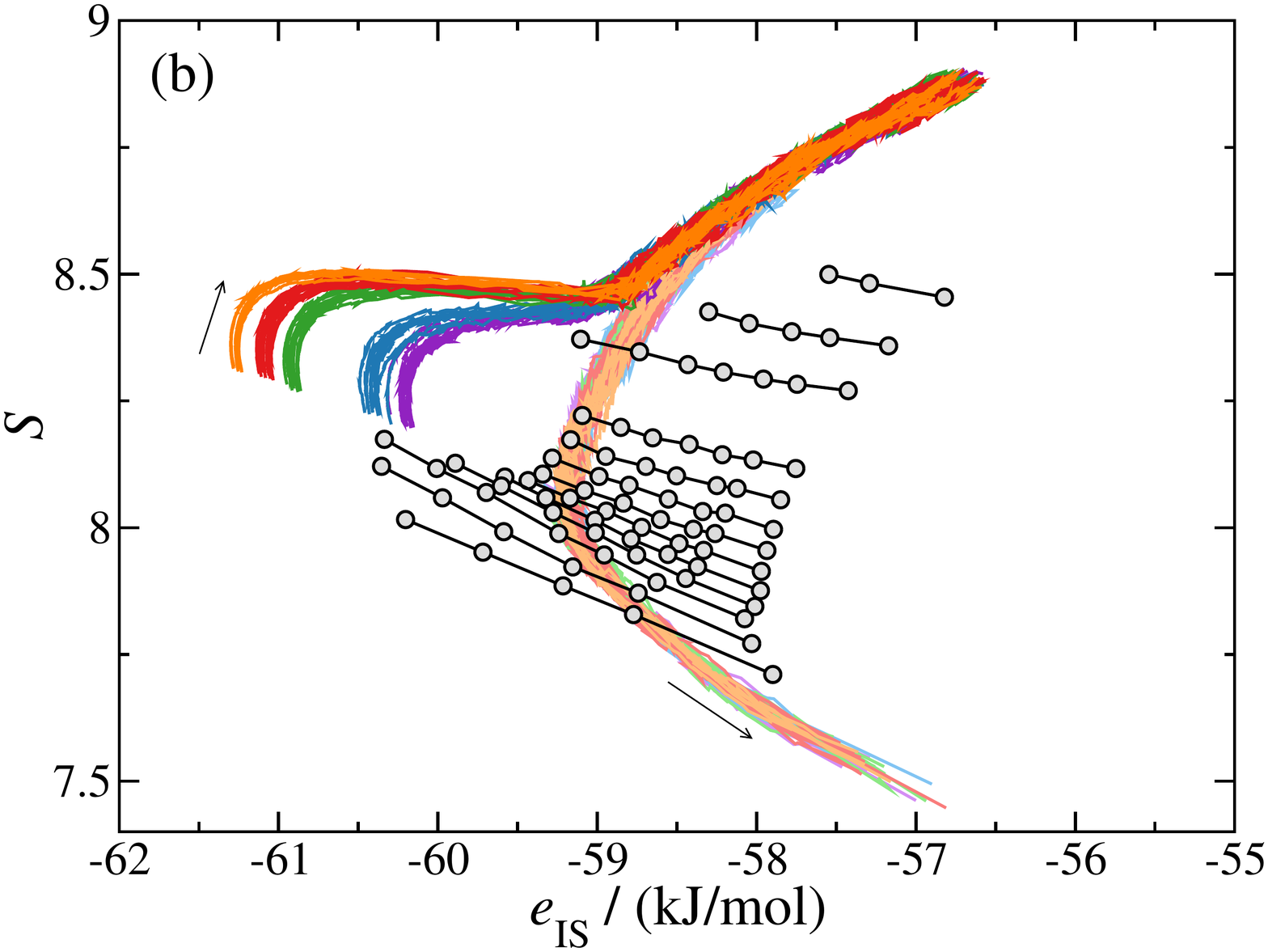}
\caption{
(a) IS Pressure, $P_\text{IS}$, and (b) shape function, $\mathcal S$, as a function of the IS energy sampled by the system during the pressure-induced LDA$\rightleftarrows$HDA transformation (solid lines) at 80~K.
Sharp and light colors correspond, respectively,  to the LDA$\rightarrow$HDA and HDA$\rightarrow$LDA transformations.
Data is taken from Fig.~10 of the main manuscript.
LDA is prepared by isobaric cooling at $P=0.1$~MPa using using different cooling rates  $q_\text{c}=0.01$--100~K/ns.
The compression/decompression rate is $q_\text{P}=10$~MPa/ns.
The circles represent the $P_\text{IS}(e_\text{IS})$ and $\mathcal{S}(e_\text{IS})$ for the stable/metastable equilibrium liquid (see Fig.~\ref{fig:delta-tdep}).
The PEL regions sampled by the equilibrium liquid and the amorphous ices are, in general, different.
}
\label{fig:delta-qdep}
\end{center}
\end{figure}

Figs.~\ref{fig:delta-qdep}(a) and \ref{fig:delta-qdep}(b) show $P_\text{IS}(e_\text{IS})$ and $\mathcal{S}(e_\text{IS})$ during the pressure-induced LDA$\rightleftarrows$HDA transformation at $T=80$~K and for different cooling rates $q_\text{c}$.
The values of $P_\text{IS}(e_\text{IS})$ and $\mathcal{S}(e_\text{IS})$ sampled by the liquid, shown in Fig.~\ref{fig:delta-tdep}, are also included.
We again find that the PEL regions sampled by the liquid and the amorphous ices are, in general,  different.
Specifically, the IS sampled by the system during the compression-induced LDA$\rightarrow$HDA transformation (sharp color lines) are never sampled by the liquid (circles).
In particular, the regions of the PEL sampled by the liquid and the amorphous ices during compression become increasingly different as $q_\text{c}$ decreases. 
During decompression (light color lines), the PEL regions accessible to the amorphous ices and the liquid also differ.
We note that, at low pressures (low density), the amorphous ices seem to explore the same IS accessed by the liquid in equilibrium [i.e., the light color lines overlap with the circles in Figs. ~\ref{fig:delta-qdep}(a) and (b)].
However, when this occurs, the IS of the amorphous ice and the liquid have indeed different densities and hence, are different (cf. Refs.~\onlinecite{giovambattista03-prl,giovambattista16-jcp,giovambattista17-jcp}).

We also note that, as observed in Fig.~\ref{fig:delta-tdep}, the regions explored by the system during the LDA$\rightarrow$HDA and HDA$\rightarrow$LDA transformations do not necessarily coincide at the temperature and rates explored, i.e., the observed LDA$\rightarrow$HDA transformation is not reversible. 

\newpage

\section{Compression Rate Dependence}
 
\begin{figure}[h!]
\begin{center}
\includegraphics[width=0.49\textwidth]{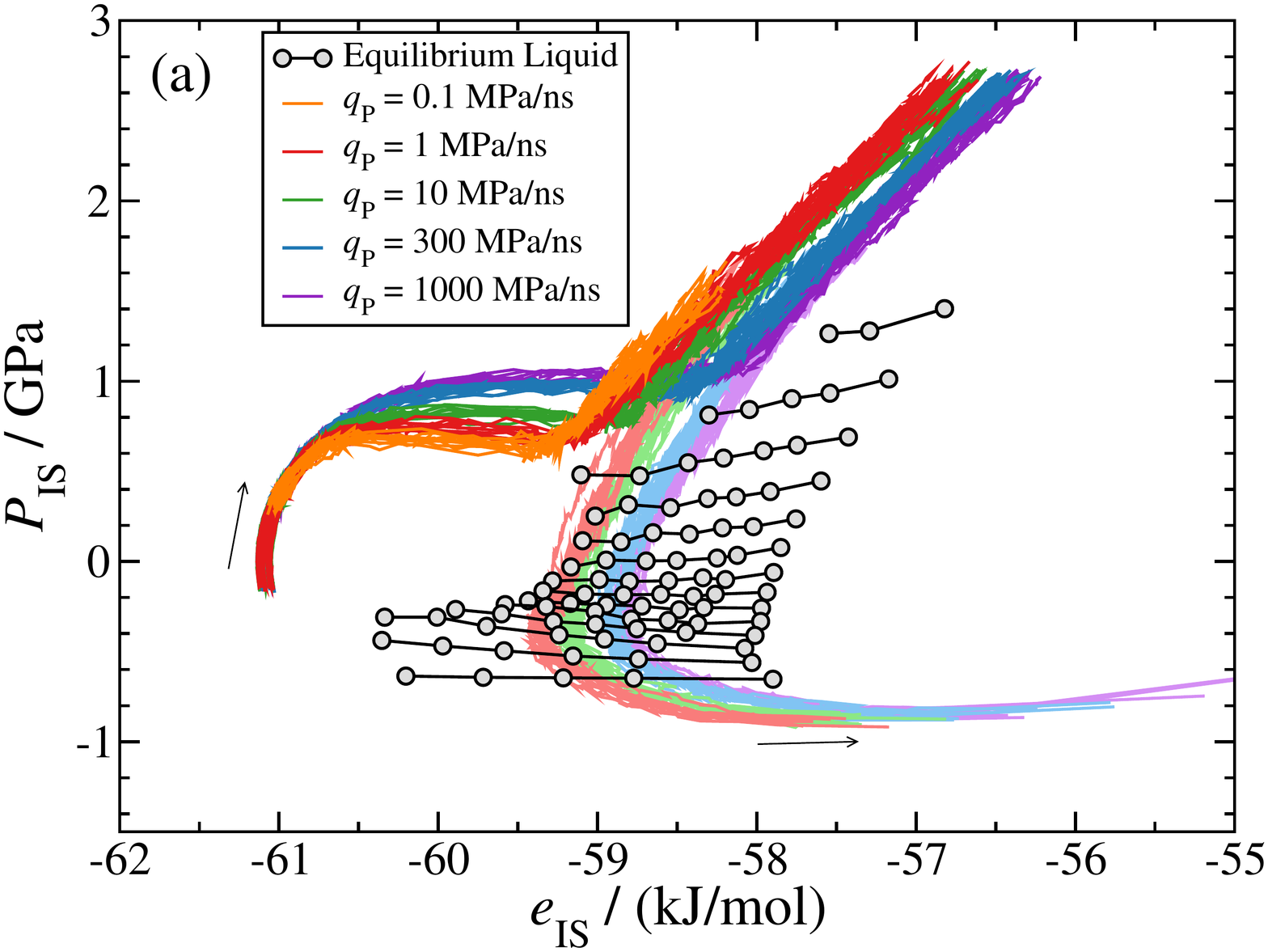}
\includegraphics[width=0.49\textwidth]{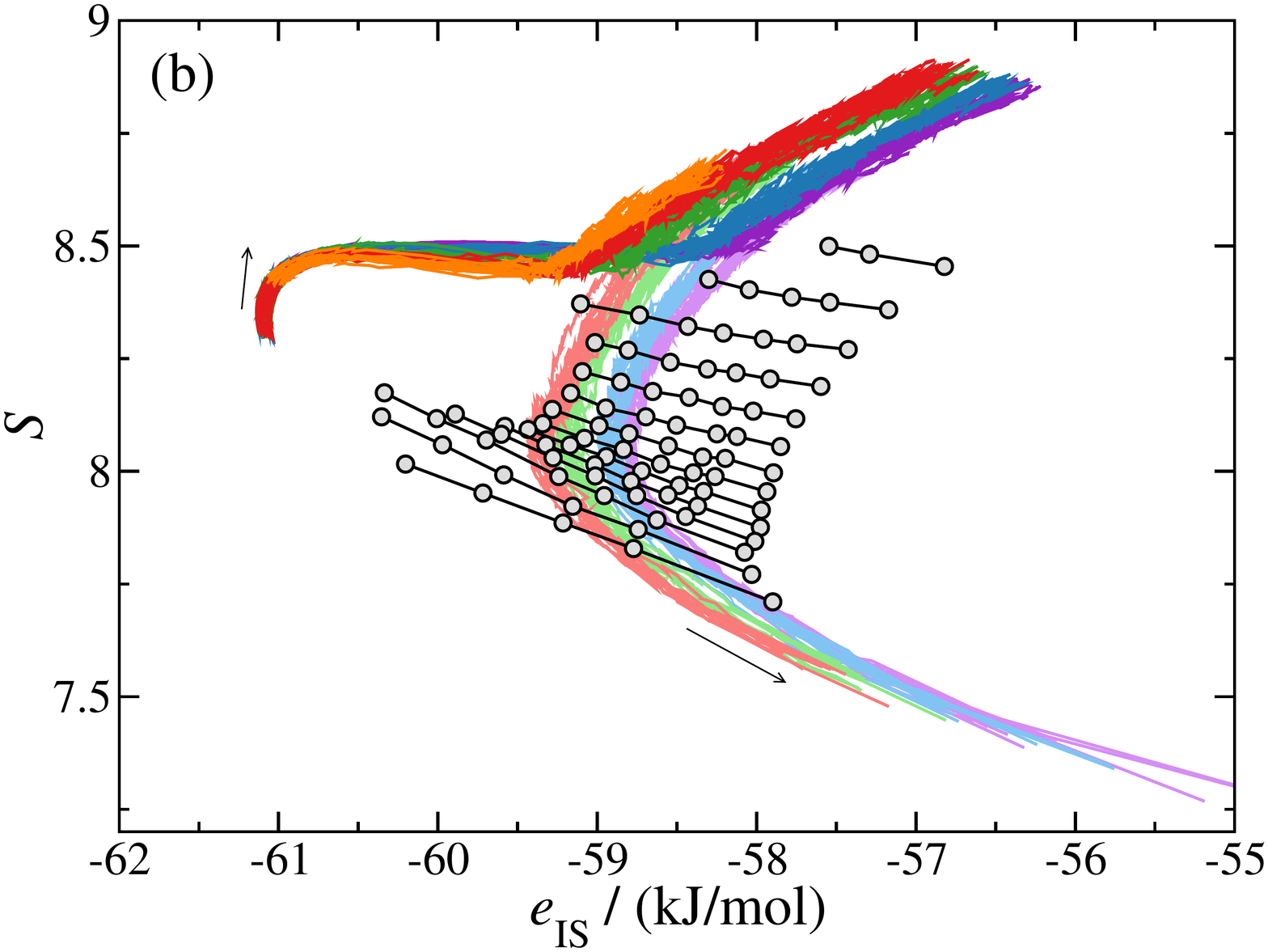}
\caption{
(a) IS Pressure, $P_\text{IS}$, and (b) shape function, $\mathcal S$, as a function of the IS energy sampled by the system during the pressure-induced LDA$\rightleftarrows$HDA transformation (solid lines) at 80~K.
Sharp and light colors correspond, respectively,  to the LDA$\rightarrow$HDA and HDA$\rightarrow$LDA transformations.
Data is taken from Fig.~14 of the main manuscript.
LDA is prepared by isobaric cooling at $P=0.1$~MPa using using a cooling rate  $q_\text{c}=0.1$~K/ns.
Different compression/decompression  rates $q_\text{P}=0.1$--$1000$~MPa/ns are used.
The circles represent the $P_\text{IS}(e_\text{IS})$ and $\mathcal{S}(e_\text{IS})$ for the stable/metastable equilibrium liquid (see Fig.~\ref{fig:delta-tdep}).
The PEL regions sampled by the equilibrium liquid and the amorphous ices are, in general, different.
}
\label{fig:delta-pidep}
\end{center}
\end{figure}

Figs.~\ref{fig:delta-pidep}(a) and \ref{fig:delta-pidep}(b)  show show $P_\text{IS}(e_\text{IS})$ and $\mathcal{S}(e_\text{IS})$ during the LDA$\rightleftarrows$HDA transformations at $T=80$~K and for different compression/decompression rates $q_\text{p}=0.1$--1000~MPa/ns.
The results are similar to those presented in Sec.~\ref{sec:qdep}.
 Briefly, the IS sampled by the liquid and the amorphous ices differ, and the transformation is not reversible, i.e., the IS sampled by the system during the LDA$\rightarrow$HDA and HDA$\rightarrow$LDA transformation are also different.

\bibliography{tip4p2005-polyam.bib}